\documentclass[11pt,letterpaper]{article}
\pdfoutput=1

\usepackage{graphicx,array}
\usepackage{graphbox}

\usepackage{enumitem}
\usepackage[dvipsnames]{xcolor}
\definecolor{darkblue}{rgb}{0,0.1,0.5}
\definecolor{darkgreen}{rgb}{0,0.5,0.2}
\definecolor{darkred}{RGB}{153,26,0}
\usepackage{ulem}
\definecolor{seablue}{rgb}{0,0.2,0.6}
\usepackage{latexsym}
\usepackage{amssymb, amsmath}
\usepackage{bm}        
\usepackage[numbers,sort&compress]{natbib}
\usepackage{bm,relsize}
\usepackage{slashed}
\usepackage{tikzfeynman}
\definecolor{viola}{RGB}{134,41,198}
\usepackage{mathrsfs}
\usepackage{hyperref} 
\hypersetup{
    colorlinks=true,       
    linkcolor=darkblue,          
    citecolor=BrickRed,        
    filecolor=magenta,      
    urlcolor=MidnightBlue           
}
\usepackage[all]{hypcap} 
\usepackage{subcaption}

\usepackage[font=small,labelfont=bf,labelsep=period]{caption}

\newcommand{\GeV}{\mathrm{GeV}}

\newcommand{\be}{\begin{equation}}
\newcommand{\ee}{\end{equation}}

 \date{\today}

\begin{document}
\thispagestyle{empty} 
\begin{flushright}

\end{flushright}
\vspace{.6cm}
\begin{center}
{\LARGE \bf Early vs late string networks from a minimal QCD Axion 
}\\
\vspace{1cm}
{
\large Marco Gorghetto$^{a,b}$, Edward Hardy$^c$, Horia Nicolaescu$^d$, Alessio Notari$^{e,f}$, Michele Redi$^g$}
\\[7mm]
 {\it \small
 $^a$Deutsches Elektronen-Synchrotron DESY, Notkestr. 85, 22607 Hamburg, Germany\\
 \vspace{.2cm}
$^b$Department of Particle Physics and Astrophysics, Weizmann Institute of Science,\\
Herzl St 234, Rehovot 761001, Israel\\
 \vspace{.2cm}
$^c$Rudolf Peierls Centre for Theoretical Physics, University of Oxford, Parks Road, Oxford OX1 3PU, UK \\
 \vspace{.2cm}
$^d$Department of Mathematical Sciences, University of Liverpool,\\
Liverpool, L69 7ZL, United Kingdom\\
 \vspace{.2cm}
$^e$ Departament de F\'isica Qu\`antica i Astrofis\'ica \& Institut de Ci\`encies del Cosmos (ICCUB), Universitat de Barcelona, Mart\'i i Franqu\`es 1, 08028 Barcelona, Spain

 \vspace{.2cm}
$^f$Galileo Galilei Institute for theoretical physics, Centro Nazionale INFN di Studi Avanzati\\
Largo Enrico Fermi 2, I-50125, Firenze, Italy\\
 \vspace{.2cm}
$^g$INFN Sezione di Firenze, Via G. Sansone 1, I-50019 Sesto Fiorentino, Italy\\
Department of Physics and Astronomy, University of Florence, Italy
 }
\end{center}

\vspace{1cm}

\centerline{\bf Abstract} 
\begin{quote}

We propose a new regime of minimal QCD axion dark matter that lies between the pre- and post-inflationary scenarios, such that the Peccei-Quinn (PQ) symmetry is restored only on sufficiently large spatial scales. This leads to a novel cosmological evolution, in which strings and domain walls re-enter the horizon and annihilate later than in the ordinary post-inflationary regime, possibly even after the QCD crossover. Such dynamics can occur if the PQ symmetry is restored by inflationary fluctuations, i.e. the Hubble parameter during inflation $H_I$ is larger than the PQ breaking scale $f_a$, but it is not thermally restored afterwards. Solving the Fokker-Planck equation, we estimate the number of inflationary e-folds required for the PQ symmetry to be, on average, restored. Moreover, we show that, in the large parts of parameter space where the radial mode is displaced from the minimum by de Sitter fluctuations, a string network forms due to the radial mode oscillating over the top of its potential after inflation. In both cases we identify order one ranges in $H_I/f_a$ and in the quartic coupling $\lambda$ of the PQ potential that lead to the late-string dynamics. In this regime  the cosmological dark matter abundance can be reproduced for axion decay constants as low as the astrophysical constraint ${\cal O}(10^8)$ GeV, corresponding to axion masses up to $10^{-2}~{\rm eV}$, and with miniclusters with masses as large as $\mathcal{O}(10)M_\odot$.

\end{quote}

\newpage
\tableofcontents

\newpage
\section{Introduction}

The cosmological dynamics of the QCD axion generally falls into one of two broad classes depending on when the Peccei-Quinn (PQ) symmetry 
is last unbroken~\cite{Peccei:1977hh,Weinberg:1977ma,Wilczek:1977pj,Kim:1979if,Shifman:1979if,Zhitnitsky:1980tq,Dine:1981rt}. In the first class,  which goes under the name of `pre-inflationary' scenario, the PQ symmetry is broken during the visible part of inflation and never subsequently restored \cite{Abbott:1982af,Dine:1982ah,Preskill:1982cy}. Conversely, in the second class, called the `post-inflationary scenario', the PQ symmetry is unbroken at some time after inflation and breaks subsequently. 

The division into the pre- and post-inflationary regimes is relevant because these possibilities could be distinguished observationally or experimentally, and they also have differing requirements on the underlying axion theory. One of the key reasons for this is that topological defects are present in the early Universe only in the post-inflationary scenario \cite{Sikivie:1982qv,Lazarides:1982tw,Vilenkin:1982ks,Vilenkin:1984ib,Bennett:1985qt,Vilenkin:1986ku,Davis:1986xc,Harari:1987ht,Shellard:1987bv,Bennett:1989yp,Battye:1993jv,Chang:1998tb}. In this case, when the PQ symmetry breaks, a network of axion strings forms and quickly approaches an attractor, scaling, solution with roughly one Hubble length of string per Hubble patch \cite{Vilenkin:1981kz,Kibble:1984hp,Albrecht:1984xv,Vachaspati:1984dz,Bennett:1987vf,Allen:1990tv}. Subsequently, when the axion mass $m_a$ reaches $m_a\simeq H$ (with $H$ the Hubble parameter), domain walls  form \cite{Sikivie:2006ni}. If the domain wall number $N_{\rm DW}$ equals unity, the strings `chop up' and quickly destroy the resulting network. If $N_{\rm DW}>1$, the domain wall network is stable and an acceptable cosmological history requires additional ingredients, e.g. an explicit PQ symmetry breaking energy bias between the $N_{\rm DW}$ vacua.\footnote{Cosmic strings are essential for a post-inflationary axion to be consistent with observations; if there were only domain walls present, infinitely large domain walls would overclose the Universe both for $N_{\rm DW}=1$ and also $N_{\rm DW}>1$ even with an energy bias.} Due to the axions produced by defects, the pre- and post-inflationary scenarios lead to almost complementary mass ranges for which an axion can account for the full observed dark matter abundance \cite{Gorghetto:2020qws,Buschmann:2021sdq}, assuming a standard cosmological history. Additionally, the spatial distribution of dark matter axions on small scales is different; relatively dense `miniclusters' form in the post-inflationary scenario but are typically absent in the pre-inflationary case \cite{Hogan:1988mp,Kolb:1993zz,Kolb:1994fi,Zurek:2006sy} (see however \cite{Arvanitaki:2019rax}). Meanwhile, pre-inflationary axions are compatible also with $N_{\rm DW}>1$, but in this case a low scale of inflation $H_I\ll  {\cal O}(10^{-5})  f_a$, where $f_a$ is the axion decay constant, is needed to evade observational constraints on isocurvature. 

The originally considered way that the post-inflationary regime can occur is if the maximal temperature of the Universe $T_{\rm max}$ after inflation~\cite{largest_temperature} is large enough, $T_{\rm max}\gtrsim f_a$. In this case the PQ symmetry is restored by finite-temperature effects and subsequently breaks through a phase transition, at some time after the end of inflation (for clarity we call this the {\it thermal post-inflationary scenario}) \cite{Weinberg:1974hy,Kibble:1976sj,Kibble:1980mv,Zurek:1985qw}.  
In this paper we instead suppose that  $H_I\gtrsim f_a \gg T_{\rm max}$. As a result, finite temperature effects do not restore the PQ symmetry after inflation. However, the PQ symmetry might be effectively restored on average on a given spatial domain  \emph{during} inflation if  inflationary fluctuations are large enough to efficiently randomise the PQ field in that region~\cite{Starobinsky:1994bd} (even if in almost all points the symmetry is locally broken), in which case the effective breaking of the PQ symmetry happens after the end of inflation when the PQ radial mode relaxes to its minimum (we refer to this as the {\it stochastic post-inflationary scenario}). { Such dynamics have been considered with the radial mode sitting in the vacuum, in which case only domain walls form, \cite{Lyth:1991ub} and with the radial mode fluctuating, which can lead to strings, \cite{Hodges:1991xs,Lyth:1992tw},  see also   \cite{Linde:1990yj,Linde:1991km}.} A naive expectation is that for a Hubble scale during inflation $H_I\gtrsim 2\pi f_a$ such restoration is inevitable and the resulting dynamics long after inflation are indistinguishable from the thermal post-inflationary case. However, {building on the analysis of \cite{Hodges:1991xs,Lyth:1992tw},} we will see that the situation is not this simple.

By studying the evolution of the PQ field during inflation, our first aim is to determine the values of $H_I/f_a$ and the quartic $\lambda$ of the PQ potential for which inflationary fluctuations do actually restore the PQ symmetry, giving rise to the stochastic post-inflationary scenario. We compare our analytic predictions to results from numerical simulations of the evolution of the PQ scalar after inflation, which provide evidence for our approach. For the same purpose, we also study  for the first time quantitatively (as far as we are aware), a mechanism noted in Ref.~\cite{Lyth:1992tw} by which oscillations of the radial mode  over the top of its potential soon after the end of inflation can amplify initially small differences between the value of the PQ scalar at different points in space, leading to a string network in parts of parameter space that would otherwise not.

Our second aim is to analyse the dynamics of the axion string network at the boundary of the parameter space where inflationary fluctuations restore the PQ symmetry, again assuming $T_{\rm max}\ll f_a$. In this case, the PQ symmetry might be restored on average only over large spatial scales, much larger than the Hubble radius at the end of inflation, while in smaller regions the symmetry is locally broken.  This corresponds to an initially underdense string network that re-enters the horizon late, possibly even after the QCD crossover, which we call the {\it late string regime}.\footnote{This effectively realizes the scenario of Ref. \cite{Redi:2022llj,Harigaya:2022pjd} where the PQ symmetry was assumed to break after the beginning of visible inflation. String networks with similar features arising as a result of different dynamics during inflation, often a phase transition, have been considered in various contexts \cite{Lazarides:1984pq,Shafi:1984tt,Vishniac:1986sk,Kofman:1986wm,Yokoyama:1989pa,Nagasawa:1991zr,Basu:1993rf,Freese:1995vp,Kamada:2012ag,Kamada:2014qta,Ringeval:2015ywa,Redi:2022llj}.
} 
Such dynamics occur for order-one ranges of $H_I/f_a$ in minimal axion theories with no additional model building, and affects both the dark matter axion mass and the properties of miniclusters. In particular, for $N_{\rm DW}=1$ the full observed dark matter abundance can be obtained for axion decay constants as small as allowed by astrophysical constraints, $f_a\gtrsim 10^8$~GeV. We also consider isocurvature perturbations in such a scenario, which might lead to important bounds although large uncertainties remain.

The structure of our work is as follows. We start in Section~\ref{s:inflation} by describing the dynamics of the PQ field during, and soon after, inflation. In Section~\ref{s:strings} we analyse how the inflationary dynamics lead to the formation of a string network and provide an analytic argument to predict when the network re-enters the horizon. This analytic analysis is confirmed by numerical simulations in Section~\ref{s:sims}. In Section~\ref{sec:iso} we discuss isocurvature perturbations in this scenario, in Section~\ref{s:landscape} we map out the landscape of the minimal axion theory we focus on and discuss the phenomenology and observational signals. Finally in Section~\ref{s:discussion} we conclude and consider directions for future work.

\section{The QCD axion  during, and immediately after, inflation } \label{s:inflation}

We consider a complex scalar field $\Phi$  described by the Lagrangian
\begin{equation} \label{eq:L}
\mathcal{L} = |\partial_\mu \Phi|^2- V \ , \qquad V=\lambda\left(|\Phi|^2-\frac{f_a^2}{2}\right)^2 , \qquad \Phi=(\phi_1+i\phi_2)/\sqrt{2} ~.
\end{equation}
The axion $a(x)$ is the phase of $\Phi(x)=\frac{\rho(x)}{\sqrt{2}}e^{ia(x)/f_a}$, while the radial mode $\rho\equiv  \sqrt{2} |\Phi|$ 
has mass $m_\rho^2= 2 \lambda f^2_a$. 
We allow the quartic to be somewhat but not too much smaller than $\mathcal{O}(1)$, e.g. $\lambda\gtrsim 10^{-4}$ (the complications for smaller $\lambda$ are discussed in Section~\ref{ss:os}). 
The PQ symmetry breaking scale is set to be $f_a$ as is the case in theories with domain wall number $N_W=1$, which give a minimal viable post-inflationary scenario. The potential $V$ in eq.~\eqref{eq:L} arises in the realization of the KSVZ model with one vector-like fermion \cite{Kim:1979if,Shifman:1979if}, however our analysis can be adapted to UV completions with other potentials and also to $N_W > 1$.\footnote{The axion potential from QCD, to be included into eq.~\eqref{eq:L}, will be irrelevant during inflation as this will occur at a high scale, with Hubble parameter of order $f_a$, in a regime far from the stochastic axion scenario \cite{Graham:2018jyp,Takahashi:2018tdu}. Note that a similar regime could also occur for domain-wall only networks~\cite{Takahashi:2020tqv,Gonzalez:2022mcx}}
The PQ symmetry breaking scale is set to be $f_a$ as is the case in theories with domain wall number $N_W=1$, which give a minimal viable post-inflationary scenario. 

\subsection{Evolution during inflation}\label{ss:FP}

We study the evolution of the PQ field during $N$ e-folds of inflation with, assumed constant, Hubble parameter $H_I$ using the stochastic formalism of inflation \cite{Starobinsky:1994bd} (see also  \cite{Jukko:2021hql,Hardwick:2017fjo}). Throughout this time, scalars with mass smaller than $\sqrt{2} H_I$ experience de Sitter fluctuations similar to thermal fluctuations with effective temperature $T_{\rm dS}=H_I/(2\pi)$ \cite{osti_4802789,Bunch:1978yq}. Once a mode exits the horizon it becomes classical with amplitude $H_I/(2\pi)$, giving a kick to average value of the field. Consequently, the field value in a coarse-grained region of size $H_I^{-3}$ undergoes a random walk, that is independent in causally disconnected regions, as well as a motion dictated by the classical potential. The combination of these two processes can be described through the Fokker-Planck (FP) equation, which gives the evolution of the probability distribution $P(\Phi)$ of the coarse-grained field sampled over a region of space that contains many Hubble volumes. 

The general form of the FP equation reads
\begin{equation}
	\frac {\partial P}{\partial t}=\frac{H_I^3}{8\pi^2} \sum_i \left[  \partial_i \partial_i P + \frac {8\pi^2} {3H_I^4} \partial_i(P \partial_i V )\right] \, ,
	\label{eq:FP}
\end{equation}
where $t$ is cosmic time and $\partial_i$ are the derivatives with respect to the fields $\phi_i$ that are light during inflation. 
In general, eq.~\eqref{eq:FP} admits a time-independent `equilibrium' solution
\begin{equation}
P_{\rm eq}(\Phi)\equiv A \, e^{\frac{- 8\pi^2 V(\Phi)}{3 H_I^4}} \, ,
\label{eq:FPdistribution}
\end{equation}
on which the spread of the field by inflationary fluctuations is balanced by the classical motion. $A$ is a normalisation constant such that $\int d\phi_1d\phi_2 P=1$.

\begin{figure}[t]
	\centering
	\includegraphics[width=0.46\linewidth]{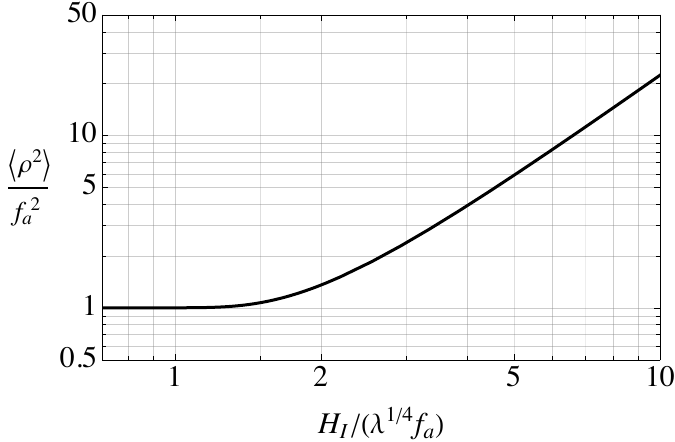} \quad
 	\includegraphics[width=0.49\linewidth]{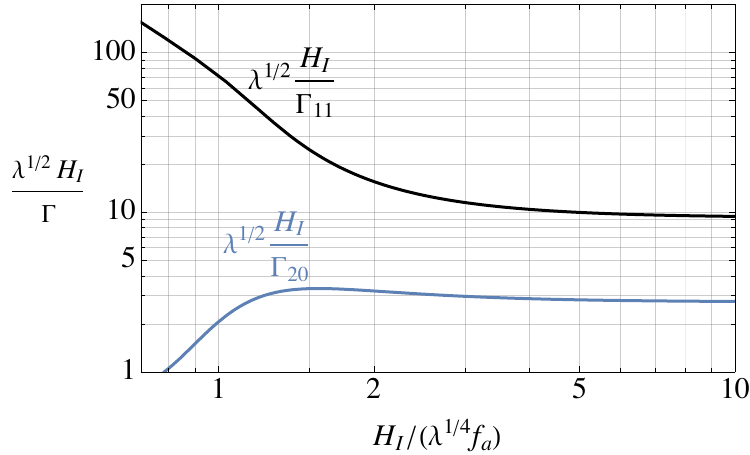} \qquad \qquad
	\caption{\label{fig:rates} {\bf Left:} The expectation value of the radial field squared $\left<\rho^2\right>/f_a^2$ on the equilibrium distribution as a function of $\alpha=H_I/(\lambda^{1/4}f_a)$. This is at the vacuum for $\alpha\ll1$ and displaced by $\alpha^2$ for $\alpha\gg1$. 
 {\bf Right:} The combination $\lambda^{1/2}H_I/\Gamma_{ij}$ corresponding to the two smallest non-vanishing eigenvalues of $O_m$, $\Gamma_{11}$ and $\Gamma_{20}$, as function of $\alpha$. The related quantity $N_{11}=H_I/\Gamma_{11}$ sets the number of e-folds on which the probability distribution of the complex scalar $\Phi$ approaches its equilibrium form during inflation.}
\end{figure}

For the Lagrangian in eq.~\eqref{eq:L}, the FP equation is two-dimensional with $\Phi=(\phi_1+i\phi_2)/\sqrt{2}$ and is valid as long as $m_\rho<\sqrt{2} H_I$ during inflation (otherwise, the formalism can only be applied to the axion, in which case string loops can only form by exponentially suppressed tunnelling processes \cite{Basu:1991ig}). On the equilibrium solution, the complex field is uniformly distributed in the angular direction and 
\begin{equation}
P_{\rm eq}(\rho)\propto e^{-\frac{2\pi^2}{3} \alpha^{-4} \left( \rho^2/f_a^2-1 \right)^{2}}~, \qquad\quad \alpha\equiv H_I/(\lambda^{1/4}f_a) \, .
\end{equation}
For $\alpha\ll 1$ the radial mode is strongly peaked in the vacuum, i.e. the expectation value  $\langle\rho^2\rangle=f_a^2$. Instead for $\alpha\gg1$ the radial mode is displaced \cite{Linde:1991km}: 
\begin{equation} 
	\langle \rho^2\rangle =  \sqrt{\frac 3 {2\pi^3 \lambda}} H_I^2  \, ,
	\label{eq:vev}
\end{equation}
with the dynamics dominated by the quartic term. In Figure~\ref{fig:rates} (left) we show $\langle\rho^2\rangle$ as a function of $\alpha$.

The asymptotic solution is approached regardless of the initial conditions. The timescale, or equivalently the number of e-foldings, on which this happens is a crucial ingredient for our work. To this aim, following \cite{Starobinsky:1994bd,Markkanen:2019kpv}, we define the reduced probability distribution
\begin{equation}
	{\tilde P}\equiv e^{\frac {4\pi^2 V}{3H_I^4} } P ~,
\end{equation}
which satisfies
\begin{equation}
 \frac {\partial \tilde{P}}{\partial t}=\frac {H_I^3}{8\pi^2} \sum_i \left[ \partial_i^2 -  \left(\left( \partial_i v\right)^2- \partial^2_iv \right)\right]\tilde{P}\,,~~~~~~ v\equiv \frac {4\pi^2}{3H_I^4} V ~.\label{eq:FP2}
\end{equation}
It is useful to express this in terms of the radial and angular modes $\rho$ and $\theta\equiv a/f_a$ to obtain,
\begin{equation} 
 \frac {\partial \tilde{P}}{\partial t}=\frac {H_I^3}{8\pi^2} \left[ \partial^2_\rho + \frac{1}{\rho}\partial_\rho   + \frac 1 {\rho^2}  \partial_\theta^2 - \left(\left(\partial_\rho v\right)^2-\frac 1 \rho \partial_\rho v - {\partial^2_\rho v}\right)\right]\tilde{P}~.
	\label{eq:tildeP}
\end{equation}
We can take advantage of the U(1) symmetry and look for solutions of the form $\tilde{P}\propto \psi_m(t,\rho) e^{im\theta}$, with $m$ integer, given the periodicity of $\theta$. This leads to
\begin{equation}
	\frac {\partial \psi_m}{\partial t}=\frac {H_I^3}{8\pi^2} \left[\partial_\rho^2+\frac 1 {\rho} \partial_\rho - \frac {m^2} { \rho^2}   - V_{\rm eff} \right]\psi_m\,,~~~~~~~~~~V_{\rm eff}\equiv 
 \left(\partial_\rho v\right)^2-\frac 1 \rho \partial_\rho v - {\partial^2_\rho v}
 ~.
\end{equation}
In this form, the right hand-side can be interpreted as the Hamiltonian of a one-dimensional quantum mechanical particle with potential $V_{\rm eff}$. In terms of $x \equiv \lambda^{1/4}\rho/H_I$ and $\alpha$, the corresponding operator is
\begin{equation}
 O_m= \frac{\sqrt{\lambda} H_I}{8\pi^2} \left[  \frac {d^2}{dx^2}+\frac 1 x \frac d {dx}- \frac {m^2} {x^2}-\frac{8\pi^2}{9}(2x^2(\pi^2(x^2-\alpha^{-2})^2-3)+3\alpha^{-2}) \right]~.
	\label{eq:rescaledFP}
\end{equation}
$O_m$ has always one zero eigenvalue, $\Gamma_{10}$, corresponding to the equilibrium solution of eq.~\eqref{eq:FP2}. The other eigenvalues (in units of $\sqrt{\lambda}H_I$) are positive and discrete, and will be denoted by $\Gamma_{nm}$ with $n>1$ integer. $\Gamma_{nm}$ and the corresponding eigenfunctions $\psi_{nm}$ can be found numerically.\footnote{For example, using Mathematica's ``NDEigensystem''.}  
The most general solution of the Schr\"oedinger-like equation~\eqref{eq:FP2} is then
\begin{equation}
\tilde P(t,x,\theta)= \sum_{nm} a_{nm} \psi_{nm}(x) \frac{e^{im\theta}}{\sqrt{2\pi}}e^{-\Gamma_{nm} t}~,
\label{eq:Ptxtheta}
\end{equation}
where the coefficients $a_{nm}$ are set by the initial conditions and we take $\psi_{nm}$ normalized in such a way that $\int_0^{\infty}dx\, x \, \psi_{nm}(x) \, \psi_{n'm}(x) =\delta_{nn'}$. Using $\psi_{10}\propto e^{-v}$ (which follows from $O_m\tilde{P}_{\rm eq}=0$ and $\tilde{P}_{\rm eq}=e^{-v}$), the probability distribution is found,
\begin{equation}
P(t,x,\theta)= P_{\rm eq}(x)+\sum_{(n,m)\ne (1,0)} a_{nm}e^{-v(x)} \psi_{nm}(x) \frac{e^{im\theta}}{\sqrt{2\pi}}e^{-\Gamma_{nm} t} ~.
\label{eq:solFP}
\end{equation}
This shows that $\Gamma_{nm}$ determine the evolution of $P$, which tends to $P_{\rm eq}$ asymptotically.  
The late-time behavior is regulated by the smallest non-zero eigenvalue, corresponding to $(n,m)=(1,1)$. In particular, $1/\Gamma_{11}$  measures the cosmic time on which  equilibrium is approached, i.e. this happens after a `critical' number of e-folds of expansion
\begin{equation}
N_{11}\equiv \frac{H_I}{\Gamma_{11}}~.
\label{N11}\end{equation}

Figure~\ref{fig:rates} (right) shows $\sqrt{\lambda}H_I/\Gamma_{nm}$ for the first two non-vanishing eigenvalues as a function of $\alpha$. 
The smallest eigenvalue in units of $H_I$ increases substantially as $\alpha$ increases  for $\alpha\lesssim 3$ (with $\lambda$ fixed), corresponding to the fact that quantum fluctuations become larger so equilibrium is reached faster. In the limit $\alpha\lesssim 1$,  $\lambda^{1/2}H_I/\Gamma_{11}\simeq (H_I/(9\lambda^{1/4} f_a) )^{-2}$ so $\Gamma_{11}$ is independent of $\lambda$. 
Meanwhile, for $\alpha\gg1$ the lowest eigenvalues are
\begin{equation}
\Gamma_{11}= 0.108 \sqrt{\lambda} H_I \,,~~~~~~~ \Gamma_{20}= 0.366 \sqrt{\lambda} H_I \,, ~~~~~~~ \Gamma_{21}= 0.611 \sqrt{\lambda} H_I\,
\label{eq:ratesHggf}
\end{equation}
and in this limit
\begin{equation}
	N_{11} \simeq \frac {10}{\sqrt{\lambda}}  ~,   \quad\qquad {\rm for}  \ \ \alpha\gg1~.
\label{eq:estimateN}	
\end{equation}
Intuitively, for $\alpha\gg 1$ the variance of $P$ initially grows as $\sigma^2 = N H_I^2/(2\pi)^2$ so the equilibrium distribution in eq.~\eqref{eq:vev} takes $N \propto 1/\sqrt{\lambda}$ e-folds to be reached, independent of $H_I/f_a$.\footnote{When $V$ is negligible a solution of eq.~\eqref{eq:FP} is a Gaussian with variance $\sigma^2=N H_I^2/(2\pi)^2$ with $N=H_It$.} Because $\Gamma_{11}$ is an increasing function of $\alpha$, eq.~\eqref{eq:estimateN} gives a lower bound for the number of e-folds to reach equilibrium, with $\rho$ light.

Importantly for what follows,  $N_{11}$ is also roughly the number of e-folds that it takes for $\Phi$ to be fully randomized; more precisely, on scales much larger than $d_{11}\equiv e^{N_{11}}/H_I$ at the end of inflation,  $a(x)$ acquires random values in the interval $[-\pi, \pi]$, while over smaller regions it has a nearly uniform value. Equivalently, two points in space that left each other's horizons fewer than $N_{11}$ e-folds before the end of inflation have correlated $\Phi$ and points separated by more than $N_{11}$ e-folds are uncorrelated (we will show this more precisely in Section~\ref{ss:corr}). Consequently, the axion field has non-trivial topology on scales larger than $d_{11}$. In particular if $N_{11}\gtrsim 50-60$ the field is nearly constant over our present Hubble patch. Meanwhile, in the opposite regime, at the end of inflation the field was inhomogeneous over what is now the observable universe, but it can still be approximately constant over smaller macroscopic regions. In the first case the PQ symmetry is effectively always broken in our universe while in the second case we can think that the symmetry is restored on average at scales larger than $d_{11}$. From eq.~\eqref{eq:estimateN} the latter scenario is realized for $\lambda \gtrsim 0.05$, when $\alpha\gg1$. 

\begin{figure}[t]
	\centering
	\includegraphics[align=c,width=0.54\linewidth]{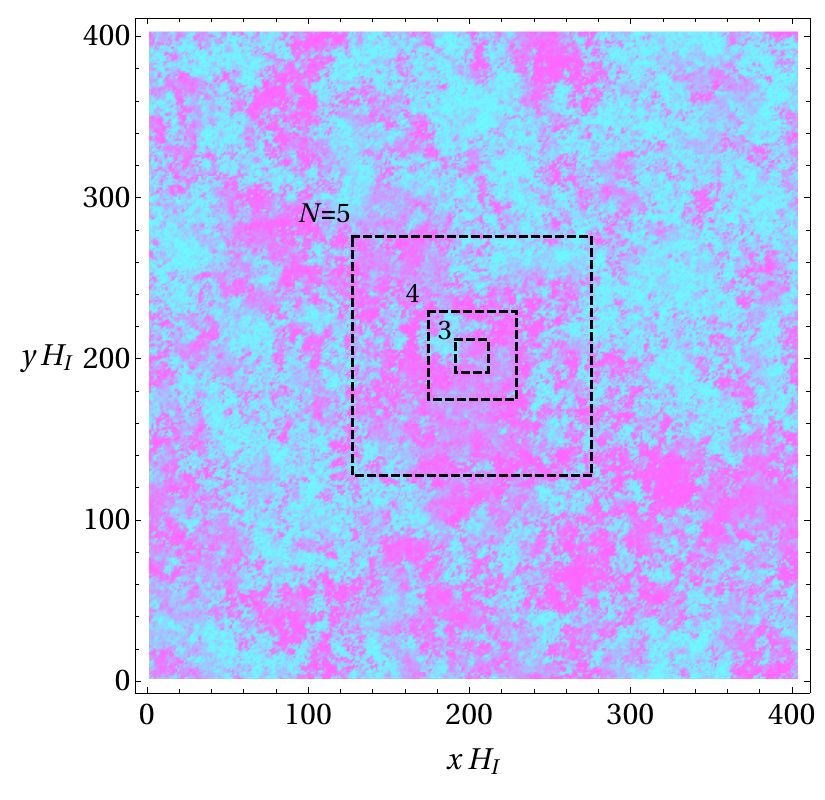} \quad
   \includegraphics[align=c,width=0.09\linewidth]{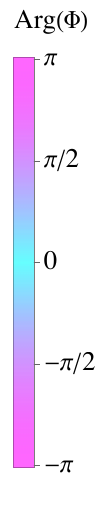}
	\caption{The axion field $a(x)/f_a\equiv {\rm Arg}(\Phi)$ over a spatial slice after 6 e-folds of inflation, with $H_I/f_a=\pi$ and the field initially in the vacuum $\Phi_0=f_a/\sqrt{2}$. This is obtained by neglecting the effect of the PQ 
 potential during inflation, i.e. we take the real and imaginary parts of $\Phi$ as Gaussian random fields with flat power spectra (this is a good approximation if $\lambda\ll1$ but not accurate for large $\lambda$, in which case $\Phi$ is non-Gaussian and the power spectra have a non-zero slope, but such an approximation is sufficient for illustrating the broad picture). Over regions separated by one e-fold of inflation, the angular field is nearly homogeneous, however over regions separated by $\simeq 4$ or more e-folds the angular field is random. Consequently, after inflation we expect the resulting string network to reach approximately one string per Hubble patch once regions that have size $e^4/H_I$ at the end of inflation have re-entered the horizon.}
\label{fig:sliceF}
\end{figure}

To illustrate the possibility of symmetry restoration only on average at sufficiently large scales, in Figure~\ref{fig:sliceF} we show $a(x)/f_a$ across a two-dimensional spatial slice  at the end of $6$ e-folds of inflation. The field is generated neglecting $V$ (sufficient for our present purposes), so $\phi_1$ and $\phi_2$ fluctuations are added to an initial homogeneous field value, arbitrarily set to $\phi_1=f_a$ and $\phi_2=0$, as Gaussian fluctuations with a flat power spectra of amplitude $H_I/(2 \pi)$ (i.e. root mean square fluctuations in the region of $\sqrt{6} H_I/(2 \pi)$). This reproduces the expectation from inflation with each point corresponding to a region of size $H_I^{-3}$ at the end of inflation. We use $H_I/f_a=\pi$, which results in the radial mode spreading over the top of the potential only after a few e-folds. Consequently, $a(x)$ is approximately homogeneous over regions corresponding to $\simeq3$ e-folds of inflation, however over larger regions it has substantial inhomogeneities, and the PQ symmetry is effectively restored. Anticipating our analysis in Section~\ref{s:strings}, in this example we expect that the string network that forms after inflation will reach one string per Hubble patch when the modes that left the horizon $\simeq 4$ e-folds before the end of inflation have re-entered the horizon.

\begin{figure}[t]
	\centering
 \vspace*{-0.5cm}
	\includegraphics[width=0.89\linewidth]{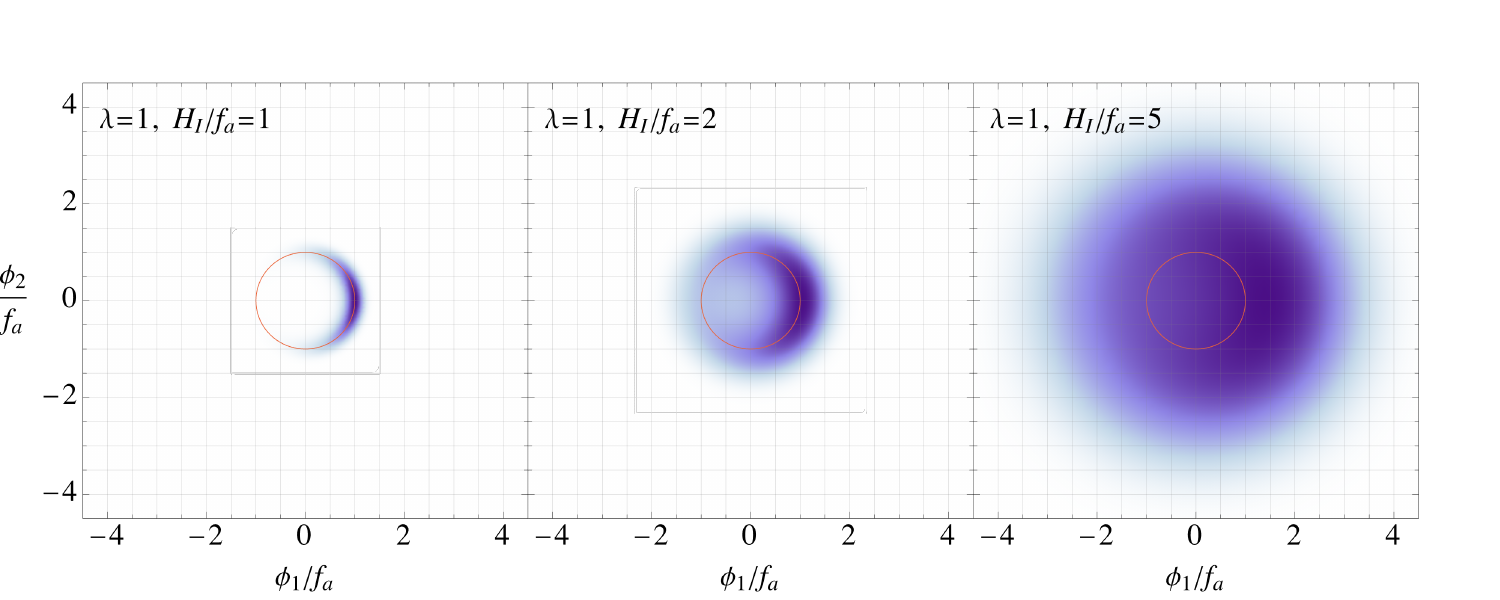}
  \vspace*{-0.5cm} \\
 	\includegraphics[width=0.915\linewidth]{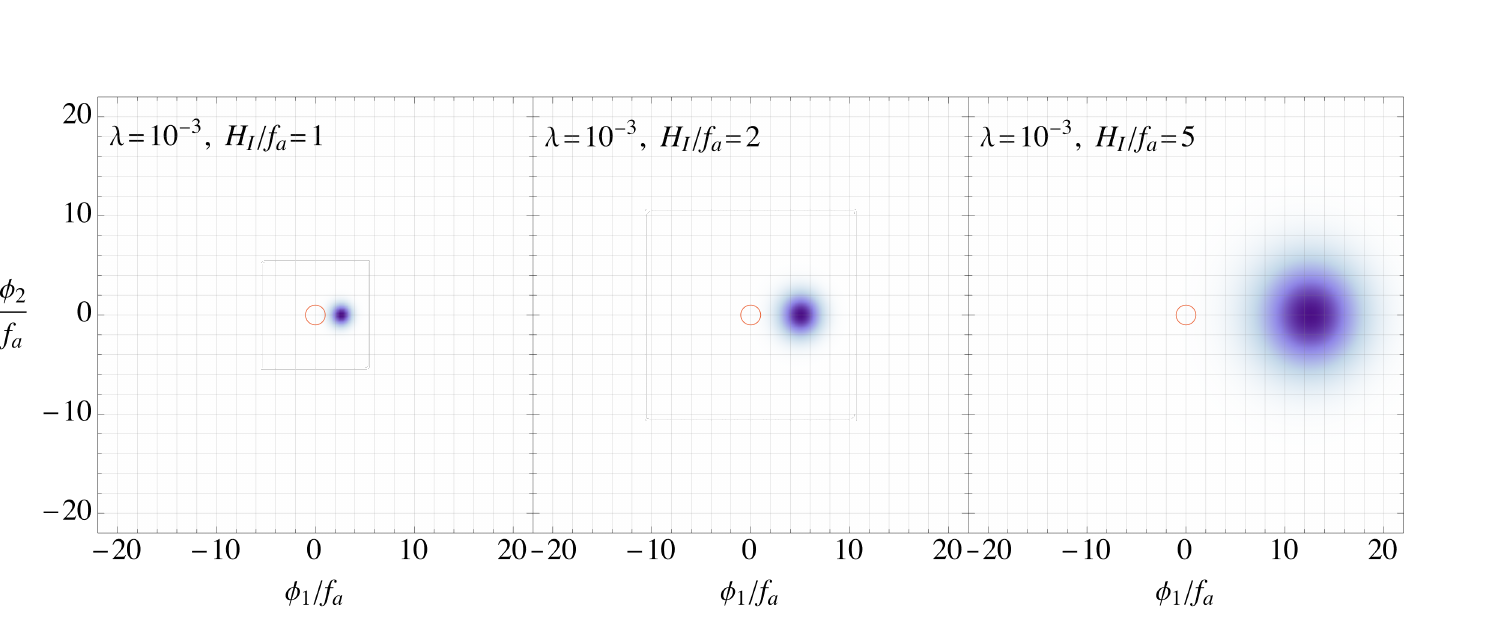}
    \vspace*{-0.125cm}
	\caption{The probability distribution $P(\Phi)$ after $N=25$ e-folds of inflation, starting from the initial condition $P=\delta(\phi_1-\sqrt{2} \Phi_0) \delta(\phi_2)$ setting $\Phi_0=\sqrt{\left<\rho^2\right>/2}$, with $\sqrt{\left<\rho^2\right>}$ the expectation value on the equilibrium distribution, shown in Figure~\ref{fig:rates}. We show this distribution for different Hubble scales during inflation $H_I$ and, in the upper and lower panels respectively, $\lambda=1$ and $\lambda=10^{-3}$. At a fixed $\lambda$, $H_I/f_a$ controls both the initial field displacement and the spread of the probability distribution. Red circles indicates the vacuum $|\Phi|=f_a/\sqrt{2}$.
 }
\label{fig:FP-evolution}
\end{figure}

In the rest of the paper we assume that, over a causal patch $N$ e-folds before the end of inflation, $\Phi$ has a value $\Phi_0$ typical of the equilibrium distribution, i.e. $P=\delta(\phi_1-\sqrt{2} \Phi_0) \delta(\phi_2)$. 
For $\alpha\ll1$ this is reasonable, while for $\alpha\gg1$ this assumes that equilibrium is reached before visible inflation. In anticipation of what follows, in Figure~\ref{fig:FP-evolution} we show the probability distribution  after $N=25$ inflationary e-folds for different $\alpha=H_I/(\lambda^{1/4}f_a)$ and $\lambda$, starting from $P=\delta(\phi_1-\sqrt{2} \Phi_0) \delta(\phi_2)$, with $\Phi_0$ given by the standard deviation on the equilibrium distribution, $\Phi_0=\sqrt{\left<\rho^2\right>/2}$. These results can be thought as the field distribution over a Hubble patch that re-enters the horizon at around the QCD crossover (similar plots focusing on $\alpha\simeq 1$ were given in Ref.~\cite{Lyth:1992tw}). For $\lambda=1$ and $H_I=f_a$, $\alpha=1$ and $N_{11}\simeq 72$: the radial mode is strongly peaked in the vacuum and the equilibrium distribution is not reached in 25 e-folds, with the field spreading only part way around the minimum of its potential; for $\lambda=1$ and $H_I=2f_a$, $\alpha=2$ and $N_{11}\simeq 15$ so after 25 e-folds $P$ is part-way to its equilibrium distribution, on which it is non-zero at $\Phi=0$ with $\left<\rho^2\right>\simeq f_a^2$ still; and for $H_I=5f_a$, $\alpha=5$ and $N_{11}=9$ so the equilibrium distribution is approximately reached and on this $P$ is unsuppressed at $\Phi=0$ and the expectation value $\left<\rho^2\right>\gtrsim f_a^2$ is slightly displaced. Meanwhile, for $\lambda=10^{-3}$, $\alpha\gg 1$ and $N_{11}\simeq 300$ for all $H_I$ and $\Phi\simeq 0$ is not explored in 25 e-folds of inflation in any of these cases because for larger $H_I$ the initial displacement gets larger at the same rate as the spread of the field: given the initially sharply peaked distribution, $P$ is approximately Gaussian with variance $ 25 H_I^2/(4\pi^2)\simeq 0.6 H_I^2$, with the radial mode initially displaced from the origin by $\left<\rho^2\right>\simeq 7H_I^2$.

\subsection{Evolution soon after inflation}\label{ss:overshoot}

We now consider the dynamics of $\Phi$ after the end of inflation when reheating begins. If $N_{11}\gg1$, $\Phi$ is roughly constant over several causal patches, so to a good approximation we initially neglect the spatial gradients and analyse the evolution patch by patch.\footnote{We also neglect any initial velocity of the field that is small due to inflationary evolution.}
If $\rho\lesssim f_a$ (e.g. for $H_I/f_a\lesssim2$ and $\lambda=1$, see Figure~\ref{fig:FP-evolution}), then the evolution is straightforward: in most patches the field simply rolls along  the radial direction into the nearest minimum of $V$ (except, as we discuss later, for the possibly few patches starting close to $\rho=0$, that will become the cores of strings). 
Conversely if $\alpha=H_I/(\lambda^{1/4}f_a)\gg1$ the radial mode takes values $\gg f_a$.  
In this case the field can oscillate along the radial direction over the top of its potential $V$ many times before settling down into the minimum, potentially dramatically affecting the field spatial distribution.

\begin{figure}[t]
	\centering
	\includegraphics[width=0.70\linewidth
 ]{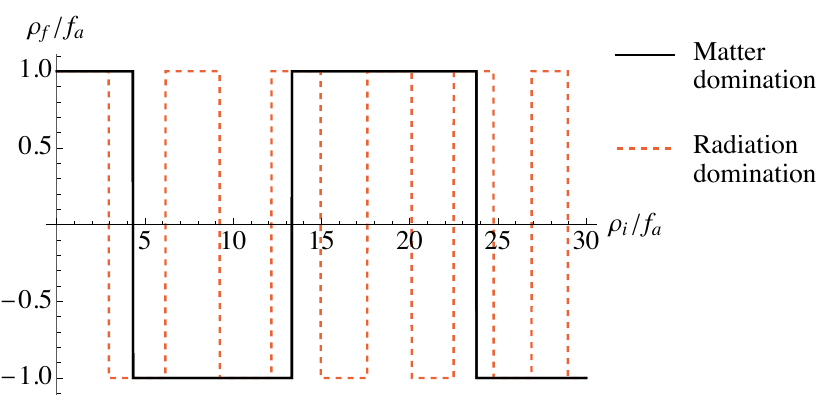}
	\caption{  Solid black line: final value of the radial field, $\rho_f$, as a function of the  value at the end of inflation, $\rho_i$, assuming a matter dominated cosmological background during the time when the radial mode relaxes to the minimum of its potential. Negative values here mean that the field ends up on the opposite side of the potential (e.g. $\theta=\pi$ when starting at $\theta=0$). Dashed red line: results obtained assuming a background of radiation domination.}
	\label{fig:phif}
\end{figure}

We focus on a simple scenario where inflation ends with the inflaton oscillating around the minimum of its potential, which we assume is quadratic so the Universe is matter dominated at this time with scale factor $R\propto t^{2/3}$. We also assume that the inflaton decay rate is sufficiently small that the maximal temperature $T_{\rm max}<f_a$; that non-perturbative processes, e.g. preheating, do not occur; and an instantaneous transition from inflation to matter domination, so the Hubble parameter is $H_I$ at the start of this era. 
With these assumptions, for the $H_I/f_a$, $\lambda$ of interest, the radial mode settles down to the minimum of its potential while the Universe is still in matter domination. During these dynamics, in each patch the equation of motion is approximated by
\begin{equation} \label{eq:eomovershoot}
\ddot{\rho}+ \frac 2 t \dot{\rho} + \lambda (\rho^2-f_a^2) \rho=0 ~.
\end{equation}
$\lambda$ can be eliminated by defining $ \sqrt{\lambda} t=\tau$, and 
reheating starts at $\tau_0=2\sqrt{\lambda}/(3H_I)$. The motion is one dimensional if we simply allow for negative values of $\rho$ (otherwise $\theta$ changes from 0 to $\pi$ and the field crosses the top of the potential).
The final value $\rho_f$ at the end of the field's oscillations is plotted in Figure~\ref{fig:phif} as a function of its initial value at the end of inflation $\rho_i$.\footnote{This function depends on $H/(\lambda^{1/2}f_a)$, as can be seen from eq.~\eqref{eq:eomovershoot} and the rescaling of $t$, however the dependence is weak and will not affect our main results in Figure~\ref{fig:Ninf}. The result of Figure~\ref{fig:phif} is shown for $H_I/(\lambda^{1/2}f_a)= 30$.} We also plot the function that one would obtain assuming a background radiation domination, in order to show that the results depend critically on this assumption. The difference is due to the larger friction term during matter domination.

The oscillation of $\rho$, which we call \emph{overshooting}, is important because it can at least partially randomise $\Phi$ even if this was close to uniform over $e^N$ patches immediately at the end of inflation, as e.g. in the bottom panel of Figure~\ref{fig:FP-evolution} for $N=25$. To see this, consider the values of the PQ field at two points in space, $x$ and $0$, such that at the end of inflation $\arg\left(\Phi(x)\right)\simeq \arg\left(\Phi(0)\right)$ and $|\Phi(x)|-|\Phi(0)|\ll |\Phi(0)|$. In the absence of overshooting, the value of $\Phi$ at these points would remain strongly correlated. However, if the small difference between $|\Phi(x)|$ and $|\Phi(0)|$ is enough that the radial mode settles down on opposite sides of the potential, i.e. if one of the steps in Figure~\ref{fig:phif} is crossed, then $\arg(\Phi(0))$ and $\arg(\Phi(x))$ will differ by $\pi$ once the radial mode has reached its minimum, effectively amplifying the initially small relative difference in the field values.  We will see that such overshooting can lead to the formation of strings: this is because of special initial field values -- the boundary points  -- where the field ends up exactly on the top of the potential, which will become string cores (this possibility has previously been noted in Ref.~\cite{Lyth:1992tw}).

\begin{figure}[t]
	\centering
	\includegraphics[align=c,width=0.39\linewidth]{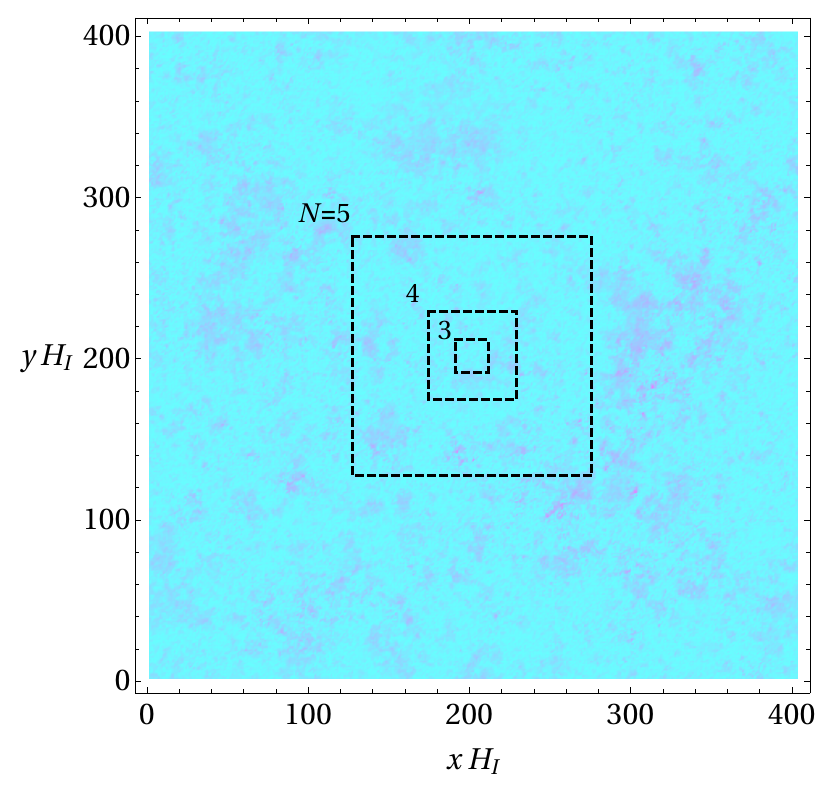}		 
 \includegraphics[align=c,width=0.10\linewidth]{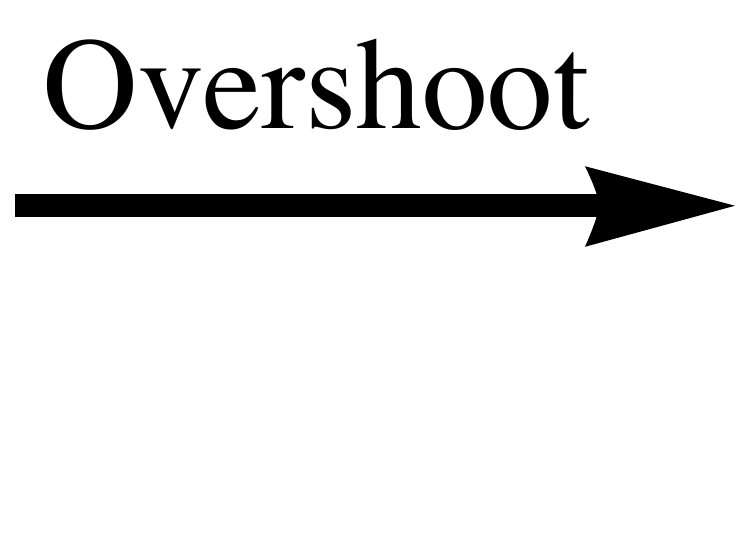}
 \includegraphics[align=c,width=0.39\linewidth]{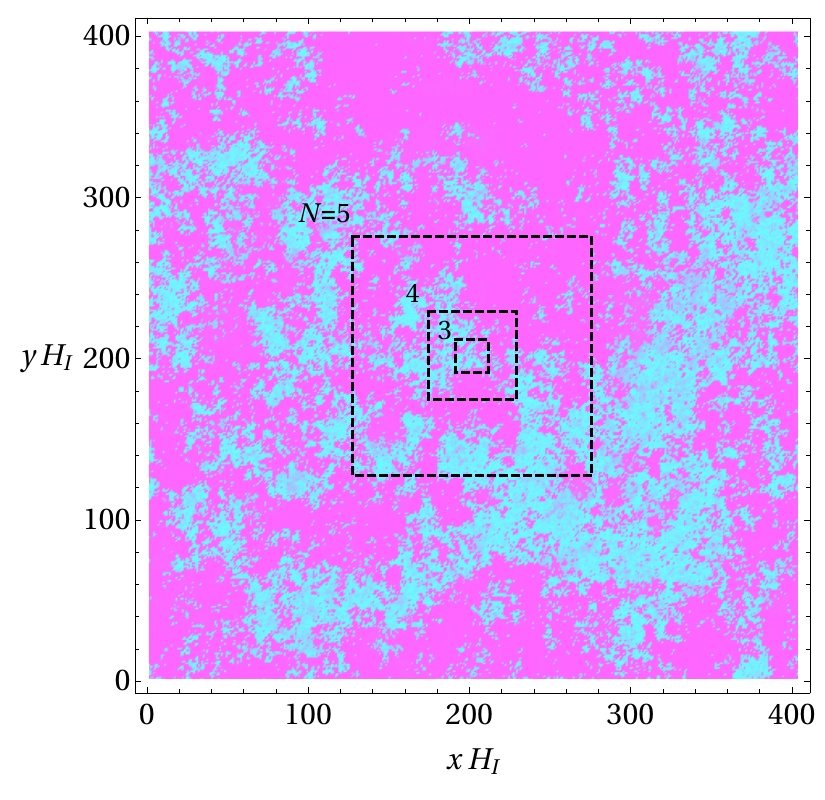}\quad
  \includegraphics[align=c,width=0.07\linewidth]{figs/sliceL.pdf}
 	\caption{ {\bf Left:} Example of $a/f_a\equiv{\rm Arg}(\Phi)$ over a spatial slice at the end of 6 e-folds of inflation, similar to Figure~\ref{fig:sliceF}, except for $H_I/f_a=\pi/2$ and with the crucial difference that the initial value of the PQ field is taken to be $\Phi_0\simeq 2.3 f_a$. We generate the fluctuations of $\phi_1$ and $\phi_2$ neglecting the effect of the potential $V$, but note that the $\Phi_0$ we choose corresponds to $\lambda=10^{-2}$ assuming $|\Phi_0|$ given by the expectation value in the equilibrium distribution eq.~\eqref{eq:vev} ($\lambda$ is sufficiently small that the effect of the potential during inflation is indeed negligible). Only a small portion of the field has spread over the top of the potential and there are only relatively small inhomogeneities just after inflation. {\bf Right:} The same field configuration as in the left panel, but after accounting for overshooting, i.e. once the radial mode has settled down to the minimum of its potential, by solving eq.~\eqref{eq:eomovershoot} (neglecting gradients and assuming an instantaneous transition to matter domination). The radial mode oscillates over the top of its potential and the initially relatively small inhomogeneities can change which side of its potential it finishes on. As a result, the angular field is efficiently randomised. This is efficient enough that strings form, and subsequently reach ${\cal O}(1)$ string per Hubble volume when  $\simeq 5$ e-folds of inflation re-enter the horizon.}
\label{fig:flatO}
\end{figure}

To illustrate the effect of overshooting, in Figure~\ref{fig:flatO} we show $a/f_a=\arg(\Phi)$ over a two-dimensional spatial slice for a realisation of the inflationary initial conditions. As in Figure~\ref{fig:sliceF}, we  generate $\phi_1$ and $\phi_2$ as Gaussian random fields with flat power spectra (i.e. we neglect $V$) and fluctuations corresponding to 6 e-folds of inflation. We set $H_I/f_a=\pi/2$, and unlike Figure~\ref{fig:sliceF}, we assume $\Phi_0=2.3 f_a$, corresponding to $\sqrt{\langle\rho^2\rangle}$ in eq.~\eqref{eq:vev} for $\lambda=10^{-2}$. These values are such that $\Phi$ does not fluctuate over the top of its potential during inflation. As a result, in the left panel, which shows $\arg(\Phi)$ immediately at the end of inflation, only small inhomogeneities are present. However, in the right panel we plot $\arg(\Phi)$ after the evolution to the minimum and overshooting, still assuming $\lambda=10^{-2}$. Evidently, the angular field acquires large fluctuations on spatial scales corresponding to $4$ or more e-folds of inflation. These will be sufficient for an (initially underdense) string network to form even though over the majority of space $\arg(\Phi)$ is close to either $0$ or $\pi$, with relatively few points in-between.

\section{Formation of strings}\label{s:strings}

As the radial mode settles down to the minimum throughout almost all of space, cores of strings form from points where the radial mode is at the top of its potential with non-trivial winding around this point.\footnote{We do not consider the regime in which the radial mode is very heavy but the axion wraps its fundamental domain many times, in which case the radial mode can be forced to the top of its potential nucleating strings (as can occur for vector dark matter produced by inflationary fluctuations) \cite{East:2022rsi}.} 
 We define the string density $\xi$ at a given cosmological time $t$ after the end of inflation as
\begin{equation}\label{eq:xi_def}
\xi = \lim_{{\cal V}\rightarrow\infty} \frac{\ell({\cal V}) t^2}{{\cal V}} ~,
\end{equation}
where $\ell({\cal V})$ is the length of strings in a volume ${\cal V}$, such that $\xi$ measures the string length in Hubble lengths per Hubble volumes. On the attractor scaling solution $\xi$ is order-one, up to corrections $\propto\log(m_\rho/H)$~\cite{Klaer:2017qhr,Gorghetto:2018myk,Kawasaki:2018bzv,Klaer:2019fxc,Buschmann:2021sdq}, while for an underdense network $\xi\ll 1$. Our key aim is to predict how many e-folds worth of inflationary fluctuations must re-enter the horizon,  which we refer to as $N$ e-folds of inflation re-entering the horizon, for $\xi\simeq 1$ to be reached; we say that at this time the string network has re-entered the horizon.

We expect that the value of $\xi$, at the time when $N$ e-folds of inflation have re-entered the horizon, is related to what extent $\Phi$ has spread over the top of the potential during the last $N$ e-folds of inflation. Intuitively, if $\Phi$ is completely spread over the top of the potential, as is the case if the equilibrium distribution is reached, over a given Hubble patch we expect at least one string.  Meanwhile, if the probability distribution of $\Phi$ is mostly concentrated on one side of the potential, we expect $\xi\ll 1$, with strings only in a few rare Hubble patches. In the latter case, $\xi$ will subsequently increase as more  e-folds of inflation re-enter the horizon, until $\xi\simeq 1$ is reached once the critical number of e-folds have re-entered.

In order to quantify the string density $\xi$ after $N$ e-folds of inflation have re-entered the horizon, we define the fraction $F$ of the field that has spread to the other side of the potential as
\begin{equation}\label{eq:Fdef}
	F(N
 ) \equiv \frac{1}{2} \left[1- {\rm Abs} \int_{-\infty}^\infty d \phi_1 \int_{-\infty}^{\infty} d \phi_2~{\rm sgn}\left(\phi_1 \right)~B\left(\sqrt{\phi_1^2+\phi_2^2}/f_a \right)~P(\phi_1,\phi_2,N) \right]~.
\end{equation}
Here $P$ is the solution the FP equation after $N$ inflationary e-folds 
starting from the initial condition $P(\phi_1,\phi_2)=\delta(\phi_1-\sqrt{2}|\Phi_0|)\delta(\phi_2)$.\footnote{In a cosmological setting before the final $N$ e-folds of inflation there were $\simeq 60-N$ visible e-folds of inflation. Consequently, we should account for a distribution of initial conditions. For most of our subsequent work however, it suffices to take $|\Phi_0|$ to have a single fixed value.\label{ft:dist}}  In eq.~\eqref{eq:Fdef}, $B$ accounts for overshoot and is the step-like function plotted in Figure~\ref{fig:phif} that takes values {$\pm 1$} depending on whether the radial mode finishes oscillating on the same or the opposite side to it started. $F(N)$ measures the fraction of the field on the less populated side of the potential. $F=0.5$ corresponds to the field having spread completely over the potential while $F=0$ corresponds to none of the field having crossed (see also \cite{Gelmini:1988sf,Larsson:1996sp} for related analysis in the context of biased initial conditions). Note that for $\alpha\ll 1$, the radial mode remains in the minimum of its potential 
and $F=0.5$ could be reached by fluctuations only in the angular direction that would not lead to strings, however this regime is not relevant to any $H_I/f_a$, $\lambda$ of phenomenological interest.\footnote{More precisely, on the equilibrium distribution, the probability of having points at $\Phi=0$, i.e. $P(0)\propto \exp(-\frac{2\pi^2}{3} \alpha^{-4})$, must be non-negligible, and provided $\alpha\gtrsim 1.6$ the radial mode does indeed explore the top of its potential. For smaller values, say $\alpha\lesssim 1$, it might be that strings do not form even at equilibrium (giving only domain walls as in Ref.~\cite{Lyth:1991ub}), but this is in the $N_{11}>50$ region of our parameter space anyway, see eq.~\eqref{N11} and Figure~\ref{fig:rates}.} A benefit of using the measure $F$ (as opposed to other possibilities such as $N_{11}$ defined in eq.~\eqref{N11}) is that $F$ captures also  the overshoot mechanism: 
$F\simeq 0.5$ can occur either because $P$ has spread over the top of the potential during the last $N$ e-folds of inflation (e.g. as in the top-right panel of Figure~\ref{fig:FP-evolution}) \emph{or} because the field has settled evenly on the two sides of the potential after overshooting despite being almost homogeneous at the end of inflation (e.g. in the bottom-right panel), and in both cases we expect a string network to form.  

In Section~\ref{s:sims} we will present evidence from numerical simulations that $\xi$ at the time when $N$ e-folds have re-entered the horizon is indeed related to $F(N)$, for the $N\lesssim 8$ that can be simulated. In particular, we will see that $\xi\simeq 1$ is obtained for $0.3 \lesssim F \lesssim 0.45$. We therefore postulate that $\xi\simeq 1$ is obtained when $F(N)\simeq 0.4$ even when this is reached for much larger $N \lesssim 60$, as is the case for a theory for which a string network re-enters the horizon soon before or even after the QCD crossover, up to the present day. For a given $H_I/f_a$, $\lambda$, we define $N_s$ to be the critical number of e-folds for $F$ to reach some value, which we usually choose to be $0.35$ (analogous results assuming $F=0.4$ are given in Appendix~\ref{a:stringprod} to give an indication of the uncertainties).

\subsection{Inflationary formation}

\begin{figure}[t]
	\centering
	\includegraphics[width=0.475\linewidth]{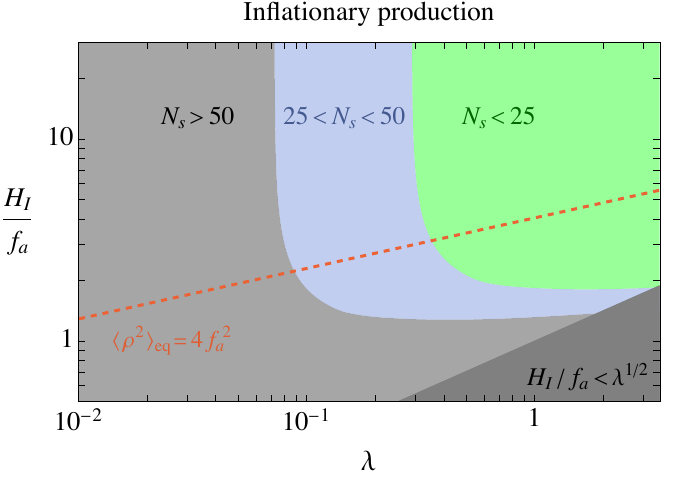} \quad
	\includegraphics[width=0.475\linewidth]{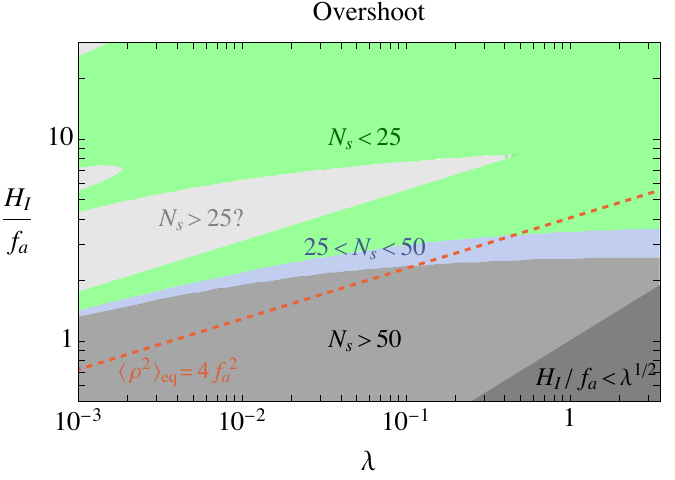}
	\caption{ {\bf Left:} Estimates of the number of e-folds of inflation $N_s$ that must re-enter the horizon for a network of strings formed from inflationary fluctuations to reach an order-one number per Hubble patch. We assume $F=0.35$ (see Appendix~\ref{a:stringprod} for the analogous plot with $F=0.4$). The $N_s<25$ region corresponds to the resulting string network re-entering the horizon before the QCD crossover. In the dark grey region in the bottom-right corner the radial mode is heavy, $m_\rho>\sqrt{2}H_I$, and our discussion does not apply (the usual isocurvature bounds on $H_I/f_a$ apply here).  
 {\bf Right:} Analogous plot but for string formation via overshoot, again with $F=0.35$. The detailed shape of the region labelled ``$N_s>25?$" is particularly sensitive to the assumed value of $\Phi$ before the visible era of inflation, the assumed critical $F$ and the cosmological history soon after the end of inflation, so this is especially uncertain. Note that in reality both production mechanisms apply, and the split here is to show the parameter space in which each is dominant. In general, one would compute $F(N)$ defined in eq.~\eqref{eq:Fdef}, which incorporates both effects, as shown in Figure~\ref{fig:landscape}.}
\label{fig:Ninf}
\end{figure}
We start with `inflationary formation' of strings, occurring when topologically nontrivial configurations are  produced directly by the radial mode spreading over its potential during inflation, such that strings form immediately after the end of inflation, in a time of order $1/m_\rho$. This amounts to neglecting overshooting by setting $B=1$ in eq.~\eqref{eq:Fdef}.

To calculate the fraction $F$, we use eq.~\eqref{eq:Fdef} and solve the FP equation for $P$. 
Assuming that the field is initially at $\phi_1=\sqrt{2} |\Phi_0|$ and $\phi_2=0$, corresponding to $x_0= \lambda^{1/4}\sqrt{2} |\Phi_0|/H_I $ 
with $\Phi_0$ sampled from the equilibrium distribution, we get that the coefficients for $P$ in eq.~\eqref{eq:Ptxtheta} are
\begin{equation}
a_{nm}= \frac{1}{\sqrt{2\pi}} e^{v(x_0)} \psi_{nm}(x_0) ~.
\end{equation}
We can now ask how the probability distribution evolves in field space over time. The fraction $F$ is simply obtained by integrating $P$ over $\phi_1<0$. The $m=0$ wave-functions on the right hand side of eq.~\eqref{eq:Ptxtheta} do not contribute to $F$ in eq.~\eqref{eq:Fdef} given their orthonormality. For $F$ sufficiently close to $1/2$, the lowest eigenvalue ($m=1$) dominates and we find
\begin{equation}
F=1/2 +  a_{11} b_{11} e^{-N \Gamma_{11}/H_I}\,, ~~~~~~  b_{11}=-\sqrt{\frac 8 {\pi} }\int dx\,x e^{-v(x)} \psi_{11}(x) \, .
\end{equation}
This means that the number of inflationary e-folds required for a fraction $F$ to be reached is
\begin{equation}\label{eq:N(F)}
N(F)= \frac {H_I} {\Gamma_{11}} \log\left[\frac{2a_{11}b_{11}}{2F-1}\right] \, .
\end{equation}
Similarly to $N_{11}$, 
$N(F)$ 
is proportional to $H_I/\Gamma_{11}$ and thus is controlled by the lowest eigenvalue $\Gamma_{11}$, even though the coefficient is sensitive to the fraction $F$. Our expectation that 
$F$ determines the string density and the result $N(F)\simeq N_{11}$ for $F\simeq 0.5$, are consistent with the intuition in Section~\ref{ss:FP} that a string network reaches $\xi\approx 1$ once, approximately, $N_{11}$ e-folds of inflation have re-entered the horizon. From eq.~\eqref{eq:N(F)}, $F$ usefully 
parameterizes the plausible range of the numerical coefficient, the true values of which can only be reliably estimated through numerical simulations.\footnote{Alternatively, we could require that in a certain fraction of space the radial field is inside a circle of radius $|\rho|<r f_a$ around the origin, with $r<1$. 
Due to the angular dependence in this case only $m=0$ wave-functions contribute, so that the second wave-function with  $m=0$ would determine $N_{s}$.
This criterion however will depend both on the radius $r$ chosen and the fraction.} Using this measure we define a critical number of e-folds $N_s$ for the network to form, using the threshold $F(N_s)=0.35$, i.e. assuming that 35\% of points reaches the other side of the potential with respect to $\Phi_0$. In general it differs by $N_{11}$ by a coefficient of about 2.

In Figure~\ref{fig:Ninf} (left), we show $H_I/f_a$ and $\lambda$ for which  
the network to re-enters the horizon at the critical number of e-folds $N_s$.
We show contours for $N_s=25,50$ corresponding to a string network re-entering the horizon just before the QCD crossover and shortly before the present day respectively; using $N_s=N_{11}=H_I/\Gamma_{11}$ gives similar results. From Figure~\ref{fig:rates} (left), for $\alpha\lesssim 2$, corresponding to the region $H_I/f_a \lesssim 2\lambda^{1/4}$ below the dotted line, the {equilibrium} distribution is peaked at the radial mode's vacuum and in this regime $\Gamma_{11}/H_I$ is approximately independent of $\lambda$. Thus, the contours are horizontal at $H_I/f_a\simeq 9/\sqrt{N_s}$ and larger $H_I/f_a$ simply results in $\Phi$ reaching  equilibrium after a smaller number of e-folds. 
Conversely, for $\alpha\gg 2$,  $\lambda^{1/2}H_I/\Gamma_{11}$  
does not change as $H_I/f_a$ increases at fixed $\lambda$, because the radial mode is displaced at $\langle \rho^2\rangle\gg f_a^2$, see 
Figures~\ref{fig:rates},\ref{fig:FP-evolution}. This leads to vertical contours 
at $\lambda\simeq 0.2(N_s/25)^{-2}$. As we will see next, in this region of parameter space the prediction from inflationary fluctuations is superseded by the effects of overshoot.

\subsection{Overshoot mechanism} \label{ss:os}

If $P$ does not approach the equilibrium distribution within the visible number of e-folds of the Universe (e.g. $\lambda\lesssim 0.05$, for $H_I \gg f_a $), naively no string network would form.\footnote{See Chapter~14 of Ref.~\cite{Gorbunov:2011zzc} for a detailed  discussion.} However, as discussed in Section~\ref{ss:overshoot}, a string network might still form after inflation by the overshoot mechanism. Yet, it is not a priori clear if this could lead to the standard scaling solution, or if it inevitably produces either an underdense network or a collection of small loops that do not percolate and form a network of long strings. The results from simulations that we present subsequently show that for $0.3 \lesssim F \lesssim 0.45$, a string network with $\xi\simeq 1$ is indeed reached.

In Figure~\ref{fig:Ninf} (right) we show contours of $N_s$, again defined by $F(N_s)=0.35$ as function of $H_I/f_a$ and $\lambda$ considering only overshoot. This means that we remove the sign function from the definition of $F$ in eq.~\eqref{eq:Fdef}, so that it is only the overshoot function that results in $F\neq 0$. We use the overshoot function in Figure~\ref{fig:phif} and, as in the case of inflationary production, we set as initial condition $\Phi_0$ sampled from the equilibrium distribution. In the time-evolution of eq.~\eqref{eq:Fdef}, $P$ is approximately Gaussian in the parameter space for which overshoot is important, i.e. the equilibrium distribution is not reached during visible inflation, with variance $N(H_I/2\pi)^2$. 
Additionally, since for the relevant $H_I/f_a$ and $\lambda$ the field does not reach the asymptotic distribution during visible inflation, other initial conditions {(i.e. $\Phi_0$)} are possible but perhaps less plausible.

At a fixed $\lambda$, as $H_I/f_a$ increases the spread of $P$ over $N$ e-folds of inflation increases, and the ``steps'' {of $B$} in Figure~\ref{fig:phif} do not get much wider at larger $\Phi_0$; thus the angular mode is increasingly randomised. The contours do however show a complex shape, labelled `$N_s>25?$', which is associated to the details of the step function determining overshoot. Initial field values close to one of the steps can easily lead to strings because only a small change is needed for the radial mode to end on the other side, whereas if the initial field value is in the middle of one of the steps it is relatively harder to form strings. 
The latter leads to regions parallel to the line $H/f_a\propto \lambda^{1/4}$ (since $H/\lambda^{1/4}$ controls the equilibrium field displacement used in $\Phi_0$) that correspond to the steps of $B$. The regions end at large enough $H_I/f_a$, because at large $H_I$ the field's probability distribution is always spread enough that it does not fully fit in a single step of $B$. We stress that the shape of these complex regions depends in detail on the initial field value; the critical value of $F$, which we do not know precisely; and the details of the overshoot function assumed, which depends on the expansion history immediately after inflation (in making this plot we assume that of Figure~\ref{fig:phif}).\footnote{Additionally, in calculating $F(N)$ to make this plot, we approximate that the radial mode $\rho$ always has value $\rho_0$ $N$ efolds before the end of inflation. In reality however, over what the now the observable Universe $\rho$ has some distribution of values due to the preceding efolds of inflation. If taken into account, this would smear our some of the sharp features in the Figure. Further details can be found in Appendix~\ref{ss:sim_description}.} As a result, these regions of the plot should be taken as a indication of the values of $H_I$, $\lambda$ at which such, complex, dynamics might occur. The result of overshoot is that string can form with relatively small $N_s$ also in the region $H_I/f_a \gtrsim 2\lambda^{1/4}$, i.e. above the dotted line, where inflationary production is inefficient, see Figure~\ref{fig:landscape} (left).

Finally, we note that Ref.~\cite{Kawasaki:2013iha} finds that if the PQ field reaches values $  |\Phi|\gg 10^4 f_a$ strings can form after inflation as a result of parametric resonance (see also \cite{Kofman:1995fi,Kofman:1997yn,Tkachev:1995md,Tkachev:1998dc,Kasuya:1997ha,Kasuya:1998td,Kasuya:1999hy,Harigaya:2015hha,Ema:2017krp}). Such field values can occur with $\Phi$ roughly at its equilibrium value eq.~\eqref{eq:vev} for $\lambda\ll 10^{-2}$ (much smaller than those we consider), or it could simply be assumed that at the start of the visible era of inflation $\Phi$ took a much larger field value than eq.~\eqref{eq:vev}. Similarly to the overshoot mechanism, parametric resonance leads to strings as a result of the radial mode oscillating over the top of its potential immediately after inflation. 
However, in parametric resonance it is the small fluctuations in the direction orthogonal to $\Phi$'s large oscillations that are dominantly amplified. Such dynamics are relevant in complementary parts of parameter space to those we focus on.

\section{Comparison to simulations} \label{s:sims}

We now present results from numerical simulations of the string network. We stress that the purpose of these is to test the general picture described above and demonstrate that the fraction $F$ defined in eq.~\eqref{eq:Fdef} sets the string density of an underdense network, at least approximately. However, as we discuss shortly we cannot directly study the regime of interest, and we also modify the physical system in {some} ways that deviate from the true dynamics.

In our simulations, the field $\Phi$ is discretized on a lattice with $n^3$ grid points. {We set the field configuration in the initial conditions to reproduce that from inflation, by generating a realisation of the Langevin equation \cite{Starobinsky:1982ee}.} To do so, we follow a method described in~\cite{Lyth:1992tw}, see Appendix~\ref{app:sims}. At the end of this, each lattice point corresponds to a single region of size $H_I^{-3}$ at the end of inflation. Given that we can only simulate grids with $n^3\sim 2048^3$ lattice points, our entire simulation volume corresponds to $N\sim 8$ e-folds of inflation, much smaller than the regime $25\lesssim N\lesssim60$ of our interest. 

We subsequently evolve the field forward in time with the discretized equations of motion of the Lagrangian in eq.~\eqref{eq:L}, using a standard leap-frog algorithm.  We start our simulations at $H=m_\rho$ for all the values of $H_I$ and $\lambda$ used in generating the initial conditions. Moreover, we evolve the mass of the radial mode throughout the simulations as $m_\rho \propto 1/R(t)$ where $R(t)$ is the scale factor, by using a time dependent $\lambda$. This is known as the ``fat-string'' trick, which means the string core size grows in time so that it contains a constant number of lattice points throughout the simulation (in practice $m_\rho \Delta=1$ is sufficient to resolve the string cores where $\Delta$ is the physical lattice spacing). Although the equations of motion of fat string system differ from {those of} the physical system by order-1 amounts, many features of the two networks agree. As discussed in Appendix~\ref{app:sims}, these choices maximise the number of e-folds worth of modes from inflation that we can observe re-entering the horizon. We evolve the system in radiation domination, $H=1/(2t)$, as is appropriate to the time when the string network re-enters the horizon in the late-string regime. Such an expansion history is unlikely to be realistic at the time when radial mode relaxes to the minimum of its potential after inflation. However, our aim is to test whether, in the $\xi\ll 1$ (under-dense) regime, $\xi$ depends on $F$ and this is not expected to be sensitive to the details of the cosmological expansion (we cannot exclude that the exact value of $F$ that leads to a given $\xi$ might have some dependence, but this is beyond the precision we can aim for).

As the system is evolved, a string network automatically forms and interacts (with strings first being cleanly defined objects at $m_\rho(t)/H(t) \simeq 10$). We calculate the length of strings in the simulations, and correspondingly $\xi$ as defined in eq.~\eqref{eq:xi_def}, using a standard algorithm \cite{Fleury:2015aca}. The simulations can be run until $HL\simeq 2$, where $L$ is the physical box size, at which point finite volume systematic errors become significant.

We extract $\xi$ at regular intervals throughout each simulation. For each such ``time-shot'' we consider the number of e-folds of inflation $N$ that have re-entered the horizon at such a time. Then we calculate the corresponding value of $F$ for each set of initial conditions by numerically solving (with Mathematica's ``NDSolve") the FP equation, starting from $P(\phi_1,\phi_2)=\delta(\phi_1-\sqrt{2}|\Phi_0|) \delta(\phi_2)$, for $N$ e-folds (i.e. until $t=N/H_I$) and finally inserting the resulting probability $P$ into eq.~\eqref{eq:Fdef}, including the function $B$ that accounts for overshoot.\footnote{We could also have extracted $F$ from the initial conditions of the simulation, which would be equivalent in the limit of large statistics.} As expected, the calculated $F$ increases as each simulation progresses and more fluctuations re-enter the horizon, and for a given simulation time $F$ is larger when $H_I/f_a$ used to generate the initial conditions is larger (keeping $\lambda$ constant).

\subsection{Results}

\begin{figure}[t]
	\centering
	\includegraphics[width=0.475\linewidth]{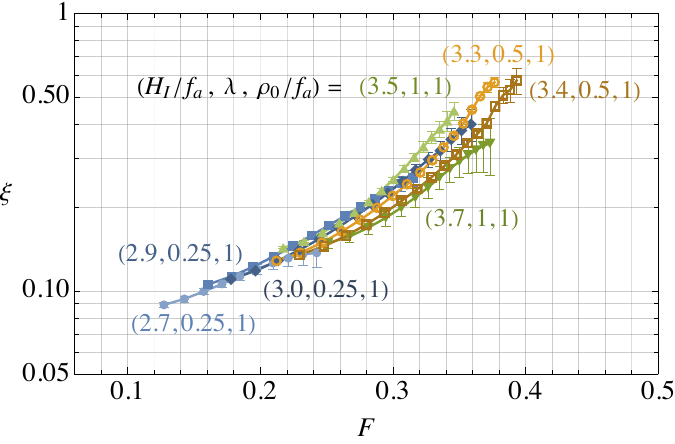}\quad
		\includegraphics[width=0.475\linewidth]{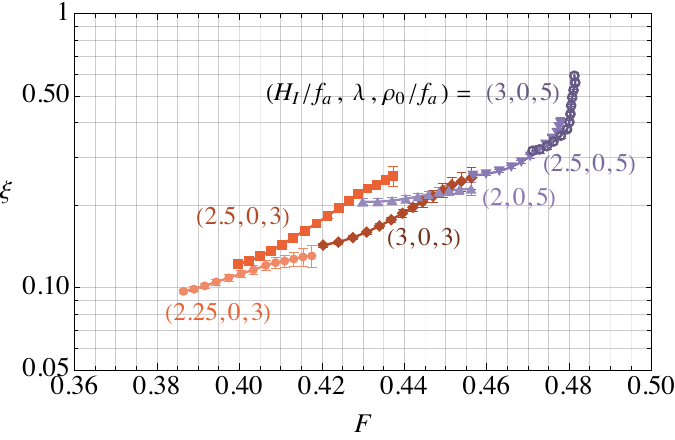}
	\caption{\label{fig:sim1} The axion string density $\xi$ in simulations starting from initial conditions corresponding to inflationary fluctuations with different value of the Hubble parameter during inflation, $H_I$, the quartic coupling, $\lambda$, and radial mode field value at the start of inflation, $\rho_0$. The results are shown as a function of $F$, defined in eq.~\eqref{eq:Fdef}, which measures the size of the inflationary fluctuations that have reentered the horizon at the time when $\xi$ is measured. On the left, data with initial conditions such that the radial mode stays mostly in the vacuum and strings form from inflationary fluctuations over the top of the potential. On the right, initial conditions such that the radial mode does not spread over the top of the potential during inflation, but is displaced from the minimum and subsequently overshoots the top of its potential leading to strings forming. Here $\lambda=0$ is the value used to generate the initial conditions, but the field was subsequently evolved with non-zero $\lambda$. In both cases, $\xi$ is approximately determined by $F$ regardless of the value of $H_I/f_a$ and $\lambda$.}
\end{figure}

We carry out simulations for a range of initial conditions (i.e. $H_I/f_a$, $\lambda$, and value of $\Phi_0$ at the start of the $\simeq 8$ simulated e-folds of inflation) leading to  
string networks that are underdense compared to the attractor, scaling, solution. The main result from these is presented in Figure~\ref{fig:sim1}, where we plot the string density $\xi$ as a function of the value of $F$ for data sets with differing initial conditions.

The results in the left panel correspond to initial conditions such that $F$ is non-zero mostly because of fluctuations of the radial mode over the top of its potential during inflation, i.e. there is very little overshooting after inflation (in particular, $\Phi_0=f_a/\sqrt{2}$). The key, non-trivial, point is that data from different sets of initial conditions, which lead to different relations between $F$ and the simulation time, approximately overlap. In other words, the values of $\xi$ at $F$ corresponding to the final time of some of the simulations match those at early simulation times in others. This suggests that $\xi$ in the underdense regime is indeed well-predicted by $F$.

Meanwhile, the results in the right panel correspond to initial conditions with relatively large initial radial mode value, $\Phi_0=\{3f_a/\sqrt{2},5f_a/\sqrt{2} \}$, such that the field is not spread over the top of the potential 
during inflation and strings can only form by the overshoot mechanism. (In such simulations, to a good approximation we neglect the potential $V$ in generating the initial conditions, i.e. we set $\lambda=0$, but 
perform the subsequent evolution after inflation with non-zero $\lambda$.) These results confirm that the radial mode overshooting indeed leads to the formation of a string network. Additionally, there is again a clear trend that larger $F$ leads to more strings although the relation is somewhat less sharp than in the left panel. We note that the values of $F$ that correspond to a particular $\xi$ are slightly larger than in the case of inflationary fluctuations over the potential.

We also note that for all the initial conditions plotted, the growth of $\xi$ as a function of Hubble parameter in simulations is $\xi= C (f_a/H)^{n}$ where the power of $n$ is roughly in the range $1/4$ to $1/3$, universally across different initial conditions and $C$ varies (see Appendix~\ref{app:sims} and Figure~\ref{fig:sim2} for details). Such a growth is consistent with an underdense string network re-entering the horizon (such an increase is distinct from the $\xi\propto \log(m_\rho/H)$ growth on the attractor solution). The value of the power $n$ is smaller than that for a network of long, non-interacting, strings which would grow as $\xi\propto 1/R^2\propto f_a/H$ in radiation domination (as in our simulations). This is expected, because as well as long strings, small loops are produced and these annihilate as they enter the horizon. We have extracted the distribution of loop lengths in our underdense networks and confirmed that such networks do indeed contain many small loops compared to the scaling distribution. Long strings are relatively rare, but it is these that ultimately fix the value of $F$ at which $\xi\simeq 1$ and scaling is reached.\footnote{Our overshoot simulations with the largest $F$ at the final simulation time (at $\log(m_\rho/H)\simeq 7$) reach $\xi\simeq 0.6$, which is not quite at the scaling value, $\xi\simeq 0.8$ at this $\log(m_\rho/H)$. However, $\xi$ is still increasing with cosmic time, and we expect scaling to be reached not much beyond the reach of such simulations.}

Of course, there are limitations to our results. For each $\lambda$, there is only a relatively small range of $H_I/f_a$ that lead to a network that is both underdense but dense enough to get reasonable statistics for $\xi$, given our available simulation volume. Related to this, as  mentioned, a key assumption of our work is that the approximate value of $F$ that leads to $\xi\simeq 0.5$ after $\simeq 4 - 7$ e-folds of inflation re-enter the horizon is similar to when $25-60$ e-folds re-enter. Despite being plausible from our results, it would be interesting to test this extrapolation in more detail, although unfortunately exponentially larger lattice sizes would be required for substantial progress.

\section{Isocurvature perturbations}
\label{sec:iso}
The inflationary fluctuations of the complex field $\Phi$ discussed in the previous sections are uncorrelated with the inflaton fluctuations. Consequently, if they ultimately lead to different axion relic abundances in different regions
, they result in dark matter isocurvature perturbations. As we review below, if a network of cosmic strings is formed at early times, reaching a scaling solution, the fluctuations of $\Phi$ on cosmological scales might be expected to be erased during the post-inflationary evolution of the field. 
According to this lore the standard post-inflationary scenario where the symmetry is restored during inflation is believed not to be affected by the corresponding 
isocurvature constraints. In contrast,
in the region of parameters that leads to the late decay of the string-wall system 
(i.e. after the QCD crossover), the initial fluctuations might survive and result in a contribution to dark matter isocurvature.

We define the axion overdensity field  $\delta_a(\vec{x}) \equiv (\rho_a(\vec{x})-\bar{\rho}_a)/\bar{\rho}_a$ where $\rho_a=\frac12 \dot{a}^2+\frac12 |\nabla{a}|^2+\frac12 m_a^2 a^2$ is the axion dark matter energy density once any strings present have been destroyed and the axion field is non-relativistic; $\bar{\rho}_a$ is its spatial average. The power spectrum $\Delta_\delta^2(k)$ of the axion overdensity field is defined by
\begin{equation}\label{eq:delta2k}
\langle\tilde{\delta}_\delta^*(\mathbf{k})\tilde{\delta}_\delta(\mathbf{k}')\rangle=\frac{2\pi^2}{k^3}\Delta^2_\delta(k)\delta^3(\mathbf{k}-\mathbf{k}') \ .
\end{equation}
where $\tilde{\delta}_a$ is the Fourier transform of $\delta_a$. We define the component of $\Delta_\delta^2$ that comprises isocurvature fluctuations to be $\Delta_{\rm iso}^2$. Meanwhile, the component of $\Delta_\delta^2$ consisting of the almost scale-invariant spectrum of curvature perturbations that is supposed to be originated by the inflaton fluctuations is $\Delta_\mathcal{R}^2(k)=A_s(k/k_{\rm CMB})^{n_s-1}$ where $A_s=2.1 \times 10^{-9}$ and $k_{\rm CMB}=0.05~{\rm Mpc}^{-1}$.

There are strong observational constraints on isocurvature fluctuations on cosmological scales.
Planck cosmic microwave background (CMB) data \cite{Planck:2018jri} bound the fraction of dark matter isocurvature fluctuations relative to the curvature perturbations. For a $\Delta_{\rm iso}^2(k)$ that is approximately constant in the interval $k=[0.002,0.1]{\rm Mpc}^{-1}$ one finds $r_{\rm iso}\equiv \Delta_{\rm iso}^2(k_{\rm CMB})/\Delta_\mathcal{R}^2(k_{\rm CMB})\lesssim 0.04$.

\subsection{Inflationary Correlations} \label{ss:corr}

We start by analysing the super-horizon correlations in the axion field immediately after the end of inflation. We first consider the regime  (following Chapter 14 of \cite{Gorbunov:2011zzc}) $H_I/f_a\gtrsim \lambda^{1/4}$ and  sufficiently small $\lambda$, such that the radial mode is displaced to field values $\gg f_a$, and that the equilibrium solution is not reached during visible inflation. As discussed, for such theories the value of the complex field $\Phi_0$ prior to the visible e-folds of inflation cannot be predicted but 
we continue to make the reasonable assumption that $\rho_0^2\simeq \left<\rho^2\right>\sim H^2_I/\sqrt{\lambda}$, where $\left<\rho^2\right>$ is the  expectation value on the equilibrium distribution, eq.~\eqref{eq:vev}. In this case after the visible e-folds of inflation,  $\Phi$ has a nearly constant classical value  $\Phi_0= \rho_0e^{i \theta_0}/\sqrt{2}$ in our Hubble patch (more generally, this is the case unless $\left|\Phi_0\right|^2 \lesssim 50 \left(H_I/(2\pi)\right)^2\ll \left<\rho^2\right>/2$).

One finds that immediately after inflation~\cite{Linde:1991km},
\begin{equation}
\left<\delta\theta^2(\mathbf{x}) \right>=  \frac{NH_I^2}{(2\pi)^2 \langle |\Phi_0|^2\rangle} ~ ,
\label{eq:isogauss0}
\end{equation}
where $\theta=\rm{Arg}(\Phi)$, $\delta \theta=\theta-\bar{\theta}$, $\bar{\theta}$ is the spatial average over the entire visible universe and $N\approx 50$ is the total number of e-folds of  visible inflation. 
If, on large scales, the correlation in eq.~\eqref{eq:isogauss0} remains unchanged from the end of inflation to the QCD scale, which neglects the effect of strings and the relaxation of $\Phi$ to the minimum via overshoot, then eq.~\eqref{eq:isogauss0} results in isocurvature perturbations in a possible contribution to the axion relic abundance produced via misalignment. Approximating $\rho_a(\mathbf{x})\propto \theta^2(\mathbf{x})$, appropriate for $\theta$ well away from $\pi$, and (for the moment) considering only misalignment production of axions, for small $\lambda$ we obtain that the isocurvature part of the power spectrum is approximately flat with an amplitude (at the end of inflation)~\cite{Beltran:2006sq,Hertzberg:2008wr}
\begin{equation}
\Delta_{\rm iso}^2=\frac {H_I^2}{\pi^2 \langle |\Phi|^2\rangle \theta_0^2} \simeq \frac {\sqrt{\lambda}}{( \pi \theta_0)^2}~,
\label{eq:isogauss}
\end{equation}
where in the last equality we used eq.~\eqref{eq:vev}. Comparing with isocurvature constraints, naively this would exclude such a scenario unless $\lambda$ is exceedingly small, $\lambda\lesssim 10^{-19}$. However, as we discuss shortly, both overshoot and the string network that can form as a result of this can modify this conclusion.

Next we study the case that $H_I/f_a$ and $\lambda$ are such that at the end of visible inflation the probability distribution of $\Phi$ is approximately given by the equilibrium solution, i.e. $N_{11}\lesssim 50$. In this regime, we can compute correlation functions using the stochastic formalism of inflation \cite{Starobinsky:1994bd}. 
We are interested in the correlation
of the phase of $\Phi$. 
Following \cite{Starobinsky:1994bd,Markkanen:2019kpv}, in the limit $\alpha=H/(\lambda^{1/4}f_a)\gg1$ one finds at the end of inflation,\footnote{Correlation functions of arbitrary 
 functions of the fields can be computed in a similar fashion. The tilt is determined by the lowest eigenvalue that contributes to the correlation function and the overlap of wave-functions.}
\begin{equation} \label{eq:coscosE}
\langle \cos \left(\theta(\mathbf{x})\right) \cos\left(\theta(0)\right)\rangle \simeq \frac {\kappa_{11}^2}{|R_eH_I\mathbf{x}|^{\frac{2\Gamma_{11}}{H_I}}}~,
\end{equation}
 where $R_e$ is the scale factor at the end of inflation and
\begin{equation}
	\kappa_{11} =
	\frac {1} {2\pi}
	\int_0^{2\pi} d\theta \cos \theta e^{i \theta}  \int_0^\infty dx \,  x \, \psi_{11}(x) \, \psi_{10}(x) \simeq 0.48  ~.
\end{equation}

As in the previous case, if we assume that the initial fluctuations in eq.~\eqref{eq:coscosE} remain unchanged until late times, they could lead to isocurvature in a misalignment-like contribution to the axion dark matter abundance. Under this assumption, we estimate $\Delta_{\rm iso}^2$ considering only misalignment production and approximating that the axion abundance $\rho_a\propto 1-\cos\theta$ (which is accurate for $\theta\ll \pi$, has the correct periodic behaviour, and is easily evaluated). This results in the overdensity power spectrum
\begin{equation}
\Delta_{\rm iso}^2(k)\simeq \kappa^2 \beta \left( \frac k {R_e H_I} \right)^{\beta} \,,~~~~~~	\beta=   \frac{2\Gamma_{11}}{H_I}~ \, .
\label{eq:isoFP}
\end{equation}
In the limit $H_I \gg f_a$ (with $\lambda\gtrsim 0.01$ so that $N_{11}\lesssim 50$, c.f. eq.~\eqref{eq:estimateN}), using eq.~\eqref{eq:ratesHggf} the spectrum tilt is $\beta=0.22\sqrt{\lambda}$. 
This can be compared with the approximate result in the mean-field approximation, described in Appendix~\ref{app:meanfield}, eq.~\eqref{eq:approxDeltacos}, which numerically gives $\alpha=0.25 \sqrt{\lambda}$. 
The isocurvature power spectrum in eq.~\eqref{eq:isoFP} at CMB scales can be written as 
\begin{equation}\label{eq:deltaiso_Nvis}
\Delta_{\rm iso}^2(k_{\rm CMB}) \sim e^{-2 N_{\rm vis}/N_{11}}  ~, \qquad N_{\rm vis}\equiv \log(R_eH_I/k_{\rm CMB}) ~ ,
\end{equation}
where $N_{\rm vis}$ is the number of e-folds of visible inflation. This implies that, for $N_{\rm vis}\simeq 50$, under the preceding assumptions the Planck constraint \cite{Planck:2018jri} would be satisfied only for $N_{11}<5$, which in turn would translate to $\lambda \gtrsim 1$ from eq.~\eqref{eq:estimateN}. However, as we have seen, for $N_{s}\lesssim 25$ a string network re-enters the horizon and reaches the scaling regime prior to the 
QCD crossover, and the effects of these strings must be taken into account. Conversely, we will argue in Section~\ref{s:landscape} that in the regime with strings reentering the horizon after the QCD crossover a contribution from misalignment as considered in this subsection will be important.

\subsection{Effects of strings on correlations}  

As mentioned, the large-scale correlations in the phase of $\Phi$ induced by inflation, see eqs.~\eqref{eq:isogauss0} and~\eqref{eq:coscosE}, are not necessarily expected to lead to significant isocurvature perturbations in the late-time dark matter relic density. In this subsection we consider how the initial fluctuations might be suppressed or erased by the effects of a string network (which might re-enter the horizon before or after the QCD crossover), and subsequently in Section~\ref{ss:isoover} we analyse the impact of overshoot.

The first possible effect occurs because a string/domain wall network emits a significant abundance of dark matter axions. If the network reaches  $\xi\simeq 1$ before the QCD crossover, axions are continuously emitted from the network to sustain the subsequent scaling regime.\footnote{Results from simulations \cite{Gorghetto:2020qws,Buschmann:2021sdq} suggest that the abundance from this source is substantially larger than the would-be abundance from misalignment.} Provided the scaling regime is indeed an attractor that does not depend on the initial conditions, the energy densities produced by strings (and, once the axion mass is relevant, domain walls) in different patches should not be correlated, as in the scenario where the PQ symmetry is restored due to thermal effects.  Moreover, in the late string regime, as we analyse in Section~\ref{s:landscape}, the string/domain wall network that re-enters the horizon after the QCD crossover gives a substantial contribution to the axion dark matter abundance; we will assume that this component carries no isocurvature (although this also deserves a deeper analysis that we leave for future work). As a result, even if there is a component of the axion abundance that arises from misalignment and retains long-distance correlations from inflation, the resulting isocurvature spectrum will be suppressed by  $\Delta_{\rm iso}^2 \propto (\rho_{\rm mis}/\rho_{\rm string})^2$, where $\rho_{\rm mis}$ and $\rho_{\rm string}$ are the axion dark matter energy densities from the `misalignment-like' production and the string/domain wall network respectively.

Secondly, if a network of strings re-enters the horizon before the QCD crossover, it is not clear whether there is even a sub-dominant contribution to the dark matter abundance from misalignment that inherits the long-distance correlations eq.~\eqref{eq:isoFP}. For $\xi\simeq 1$ at the QCD crossover, such misalignment production cannot be clearly isolated from the string/domain network, and, relatedly, $\theta$ is not necessarily constant on large scales during the evolution. 
The technical reason for this is that in the string cores $\theta$ is not defined and so $\theta$ is not a conserved quantity on large scales.\footnote{As example consider
a single string that  is initially not in the vacuum. The phase necessarily winds $[-\pi,\pi]$ around the string but the interval will not be wound uniformly.
 However, as time progresses, gradients will smooth out the field configuration driving the system to the string vacuum thus changing the value of $\theta$ around the string.} (Instead,
 in absence of strings and with the radial mode in its vacuum everywhere the zero mode decouples from the equations of motion and $\theta$  is a conserved quantity on very large scales.) As a result, a string network in scaling could erase long-distance correlations in $\theta$. 
However, to our knowledge whether this indeed happens has not been properly addressed in the literature. It could, in principle, be clarified by numerical simulations, however this is technically difficult because only $O(10)$ e-folds of evolution can be followed in the simulations, and moreover initial conditions that lead to a string network that reaches scaling quickly have a power spectrum that is already highly suppressed at large scales. We find that initial conditions with a flat power spectrum lead to an underdense network with $\xi\ll 1$ (that will only reach scaling beyond the reach of simulation), and in this case we see no erasing of large distance correlations. But this could still be consistent with the picture above, given that most Hubble patches do not yet contain a string. 
Given the importance of this question for the axion post-inflationary scenario, it would be interesting if future, larger, simulations could provide direct evidence for such effects. 

Conversely, in the late string regime, most Hubble patches do not contain a string when the Universe has temperature $T\simeq \GeV$, at which time the axion potential becomes relevant. As a result, in this case, there is a cleanly defined misalignment contribution to the axion relic abundance, which seems likely to retain the long-distance correlations from inflationary fluctuations because strings are rare. We do however note that even in this case, the small string loops that decay prior to the QCD crossover might affect the long-distance correlations even if the density of long string is not yet high enough for scaling.

\subsection{The effect of overshoot on super-horizon correlations} \label{ss:isoover}
Finally, we discuss the possibility that the relaxation of the radial mode to the vacuum -- via overshooting its potential in a significant fraction of space -- can change the large distance correlations in $\theta$. 
In particular, for distances $|\mathbf{x}|$ over which the probability distribution of $\Phi(\mathbf{x})$ immediately at the end of inflation covers many of the ``steps'' in Figure~\ref{fig:phif}, overshooting results in $\cos\theta(\mathbf{x})$ taking an effectively random value from $\{\cos \theta_0(\mathbf{x}),-\cos(\theta_0(\mathbf{x}))\}$, where $\theta_0(\mathbf{x})$ is the value before overshooting. 
Consequently, $\langle \cos \theta(\mathbf{x}) \cos\theta(0) \rangle$ can average towards zero, even if $\theta_0(0)$ and $\theta_0(\mathbf{x})$ are highly correlated.

In reality, $\cos(\theta(\mathbf{x}))$ and $\cos(\theta(0))$ are not completely decorrelated by overshooting, and the extent of the suppression by overshooting is important when confronting observational constraints. We have numerically investigated realizations of $\Phi$ with inflationary initial conditions for $H_I$, $\lambda$ such that $\left<\rho^2\right>\gg f_a^2$. Although we leave a complete study for future work, preliminary analysis suggests that overshooting does indeed suppress large distance correlations, with a suppression factor that is consistent with
\begin{equation} \label{eq:Oversup}
\frac{\langle \cos \theta(\mathbf{x}) \cos\theta(0)\rangle}{\langle \cos \theta_0(\mathbf{x}) \cos\theta_0(0)\rangle} \sim (1-2 F(N))^2~,
\end{equation}
where $N=\log(|\mathbf{x}|H_I)$ is the number of e-folds of inflation that separates $0$ and $\mathbf{x}$ (results from simulations showing a suppression can be found in Appendix~\ref{a:overshoot}). Such a dependence is plausible: as expected, the correlation vanishes if $\Phi$ is evenly distributed over both sides of the potential after overshooting (i.e. $F\simeq 0.5$), and for a correlation to remain  neither $\Phi(0)$ or $\Phi(\mathbf{x})$ can be ``randomised''. 

\section{Landscape of axion scenarios} \label{s:landscape}

We now turn to discuss phenomenology. Intuitively, theories such that a string network re-enters the horizon, i.e. reaches $\xi\simeq 1$, sufficiently early are indistinguishable from the thermal axion post-inflationary scenario (up to possible isocurvature, as just discussed). On the other hand, if a string network re-enters the horizon later, the evolution can be qualitatively different with important implications for phenomenology.

\subsection{Breaking temperature of PQ symmetry} \label{ss:small_rad}

In Section~\ref{s:strings}, we showed how the combination of inflationary fluctuations and overshoot predicts, for a given $H_I$ and $\lambda$, a critical number of e-folds $N_s$ that need to re-enter the horizon for $\xi\simeq 1$ to be reached. This corresponds to $F(N_s)$ equal to some critical value, which we assume to be $\simeq 0.35$.\footnote{In practice, in our plots we approximate $N_s ={\rm Min}[N_s^{\rm inf.}\,,N_s^{\rm o.s.}]$ from the two sources, which is a good approximation over the majority of parameter space where one or other formation mechanism dominates.} 
Defining, as in eq.~\eqref{eq:deltaiso_Nvis}, $N_{\rm vis}$ to be the total number of visible e-folds of inflation, the relevant modes exited the horizon
\begin{equation}
	N_{\rm PQ}= N_{\rm vis}- N_s~,
 \label{eq:NPQ}
\end{equation}
from the beginning of visible inflation. $N_{\rm PQ}$ has a simple relation to the temperature of the Universe when the mode re-enters the horizon. Solving $H_0 e^{N_{\rm PQ}}=R H$ in radiation domination one finds,
\begin{equation}\label{eq:TPQ}
	T_{\rm PQ}=T_0 e^{N_{\rm PQ}+4} \left(\frac {10}{g_*(T_{PQ})}\right)^{1/6} ~,
\end{equation}
where $T_0$ is the CMB photon temperature today and $g_*(T)$ is the number of relativistic degrees of freedom.  For instance, modes that renter at $T_{PQ}=1\,, 0.15\,, 0.001$ GeV exited at $N_{\rm PQ}=25\,, 23\,, 18$. Eqs. (\ref{eq:TPQ}) and (\ref{eq:NPQ}) allow to relate $N_s$ to the temperature when the string network re-enters the horizon, which is a key quantity for the phenomenology.

It is useful to compare $T_{PQ}$ with the temperature $T_\star$ when $H(T_\star)=m_a(T_\star)$, around which time the axion potential becomes cosmologically relevant and the axion starts oscillating. This reads
\begin{equation}
    T_\star=\left[\frac{3 \sqrt{10}}{\pi g_{*}^{1/2}(T_\star)} \left(\frac{\Lambda_1}{\Lambda}\right)^\frac{\alpha_m}{2} \frac{M_p}{f_a }\right]^{\frac{2}{\alpha_m +4}}\simeq 5.5\, {\rm GeV}\left[ \frac{10^8 \,{\rm GeV}}{f_a}\right]^\frac{1}{6}\left[ \frac{60}{g_{*}(T_\star)}\right]^\frac{1}{12} \, ,
\end{equation}
where $m_a(T) = m_a(T=0)(\Lambda_1/T)^{\alpha_m/2}$, with $\Lambda_1\simeq 150$ MeV, $\Lambda=75$ MeV, $\alpha_m\simeq 8$, and $m_a(0)=\Lambda^2/f_a$.

Note that $N_{\rm vis}$ depends on $H_I$ and the, uncertain, details of the cosmological expansion, e.g. soon after inflation, see eq.~\eqref{eq:deltaiso_Nvis}. For the scenarios of most interest to us, $H_I\sim f_a\sim 10^{10}~{\rm GeV}$. We also assume that after inflation the cosmological history is simply (relatively slow) reheating followed by radiation domination, in which case $N_{\rm vis}$ depends logarithmically on the reheating temperature $T_{\rm RH}$, c.f. eq.~\eqref{eq:deltaiso_Nvis}. As a reference value, we take $N_{\rm vis}=50$.

\subsection{Regimes}

Depending on the value of $N_s$ different scenarios are realized:

\begin{figure}[t]
	\centering
	\includegraphics[width=0.60\linewidth]{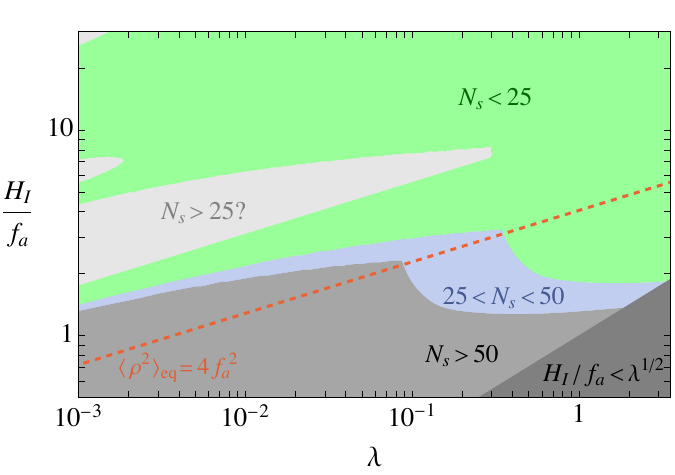} \quad
	\caption{\label{fig:landscape}  Estimate of number $N_s$ of e-folds of inflation that must re-enter the horizon for the resulting string network to reach a density of roughly one string per Hubble patch, i.e. scaling. We account for production of strings both directly from inflationary fluctuations and from the overshoot mechanism, with the criteria fraction $F(N_s)=0.35$ as defined in eq.~\eqref{eq:Fdef} (see Appendix~\ref{a:stringprod} for the analogous plot for $0.4$). For $H/f_a \lesssim2 \lambda^{1/4}$ (below the dotted red line) production of strings directly by inflationary fluctuations dominates, while in the opposite regime the overshoot mechanism is mostly responsible for the generation of the network. Above the red line that separates the two regimes the radial mode $\rho$ gets an expectation value $\gtrsim 2f_a$ during inflation and below its distribution is peaked in the vacuum. In the dark grey region the radial mode is heavier than Hubble so that the stochastic formalism for $\Phi$ is not applicable.}
\end{figure}

\subsubsection*{$N_s>N_{\rm vis}$: Broken PQ symmetry}

If $N_s>N_{\rm vis}$ then, for any practical purpose, the PQ symmetry is broken throughout the evolution of the universe realizing a pre-inflationary axion scenario. Because we assume $H_I \gtrsim f_a$, this regime is excluded by isocurvature perturbations \cite{Axenides:1983hj,Linde:1985yf,Seckel:1985tj,Lyth:1989pb,Turner:1990uz,Beltran:2006sq,Hertzberg:2008wr} (unless any of the following happens: (1) $\lambda\lesssim 10^{-19}$ from eq.~\eqref{eq:isogauss}, (2) the initial condition for $\Phi$ at the beginning of the visible inflation is very large, much bigger than the expected value from FP equilibrium distribution).

Note that if one were to consider only string production directly by inflationary fluctuations, we would obtain $N_s>N_{\rm vis}$ for any  $\lambda\lesssim 0.05$ since the PQ field is approximately constant in
our Hubble patch (at least for our assumed $\Phi_0$), see Figure~\ref{fig:Ninf} (left) and \cite{Gorbunov:2011zzc}. However, we have shown that due to the overshoot mechanism a network may still form when $\lambda$ is small, Figure~\ref{fig:Ninf} (right), so the region of small $\lambda$ can still be viable, falling into one of the two subsequent classes rather than the present one.

\subsubsection*{$N_s < N_{\rm vis}-25$: Restored PQ symmetry}

For such $N_s$, at the end of inflation $\Phi$ has fluctuations over our observable Universe that are large enough to form strings (either directly or by overshooting). Moreover, the resulting  string network re-enters the horizon  before 
the QCD crossover (because modes that reenter the horizon at around this time left the horizon $\simeq 25$ e-folds after the start of visible inflation). This regime leads to predictions that are similar to the axion post-inflationary scenario.

In this scenario the axion relic abundance receives at least a contribution from the axions emitted during the string scaling regime. This is uncertain because the result depends on the energy density spectrum of the axions emitted by the strings, and simulation results need to be extrapolated over orders of magnitude. However, the final result likely leads to an over-abundance of axions with respect to the misalignment, corresponding to requiring $f_a\lesssim 10^{10}\,\GeV$ for the axion dark matter abundance to not overclose the Universe \cite{Gorghetto:2020qws} (there is disagreement in the upper bound on $f_a$ arising from uncertainties in the form of the axion emission spectrum from strings \cite{Battye:1994au,Yamaguchi:1998gx,Hiramatsu:2010yu,Hiramatsu:2012gg,Kawasaki:2014sqa,Klaer:2017ond,Buschmann:2021sdq}). The gravitational waves (GWs) emitted by strings are likely to be unobservable in this scenario for $f_a\lesssim 10^{10}\,\GeV$~\cite{Gorghetto:2020qws}.

\subsubsection*{$N_{\rm vis}-25 <N_s < N_{\rm vis}$: Late strings }

In the region of parameter space such that $N_{\rm vis}-25 <N_s < N_{\rm vis}$, \emph{late strings}, the PQ symmetry is effectively broken over the observable Universe, but only on scales that re-enter the horizon after the QCD crossover. This results in a string network that re-enters the horizon, i.e. reaches $\xi\simeq 1$, only when the temperature of the Universe $\lesssim {\rm GeV}$ (more precisely, when $T_{PQ}<T_\star$, see eq.~\eqref{eq:TPQ}).  At such times the axion mass is relevant and the strings are attached to domain walls that form at $H=m_a$, and (assuming $N_W=1$) the string-domain wall network is immediately unstable. Such a regime can lead to very different phenomenology compared to if scaling is reached while the axion mass is negligible. The dynamics are effectively those of the meso-inflationary axion scenario where the PQ is broken after the beginning of visible inflation Ref.~\cite{Redi:2022llj}.

\vspace{0.25cm}

In Figure \ref{fig:landscape} we show the regions where the different behaviours are expected as a function $H_I/f_a$ and $\lambda$,  using the criteria $F(N_s)=0.35$ to determine $N_s$, with the full $F$ in eq.~\eqref{eq:Fdef} that incorporates both inflationary production and overshoot (results with $F=0.4$ are shown in Appendix~\ref{a:stringprod}; the difference gives a feeling of the uncertainty 
involved in our estimates). In green we show the region where the network forms at $T>$ GeV so that the standard post-inflationary axion phenomenology is expected,
where in particular axions are mostly emitted by the string network.
In the blue and dark white regions (labelled by `$25<N_s<50$' and `$N_s>25?$' respectively)  the annihilation of the string/domain wall network takes place at  $T\lesssim1$~GeV, i.e. after the axion mass is cosmologically relevant. 
In these regions the phenomenology is vastly different from the post-inflationary scenario. 
In the grey region the symmetry is always broken in the visible Hubble patch so that this corresponds to a pre-inflationary axion (ruled out by isocurvature). As in Figure~\ref{fig:Ninf}, and discussed in detail in  Section~\ref{ss:os}, the shape of the light gray region, `$N_s>25?$' is especially uncertain.

\subsection{The late string relic abundance}\label{ss:latestringrelic}

In the late string regime, $N_{\rm vis}-25 <N_s < N_{\rm vis}$, we can cleanly distinguish two different contributions to the final axion abundance: \emph{(i)} from the misalignment and \emph{(ii)} from the annihilation of the string-wall network that subsequently re-enters the horizon at $T=T_{PQ}$, related to $N_s$ via eq.~\eqref{eq:TPQ}.

The former arises when the axion mass becomes cosmologically relevant at $H\simeq m_a$, which happens before the inhomogeneities generated during inflation, and the string network, re-enter the horizon. Thus at this time it is a good approximation to consider the axion field to be homogeneous in every causal patch, except for the few patches that contain a string (where the axion instead wraps the fundamental domain $(-\pi,\pi]$). When $H=m_a$, the axion starts oscillating around the minimum of its potential, and, at least in those patches that do not contain a string and for which the value of the axion $a/f_a\neq \pi$ (which will contain a domain wall), this leads to a misalignment contribution to the axion dark matter abundance. Over different Hubble patches, the initial misalignment angle is randomly distributed in the interval $(-\pi,\pi]$, so the resulting averaged relic abundance is~\cite{Borsanyi:2016ksw}
\begin{equation}
\frac{\Omega_a^{\rm mis}}{\Omega_{\rm dm}}\simeq 0.4 \,\kappa \left[ \frac{f_a}{10^{12} \,{\rm GeV}}\right]^\frac{7}{6}\left[ \frac{60}{g_{*}(T_\star)}\right]^\frac{5}{12} \, ,
\label{eq:Omegamis}
\end{equation}
where $\Omega_{\rm dm}$ the observed dark matter abundance and $\kappa$ is a constant that depends on $\alpha_m$, with $\kappa\simeq 5$ for $\alpha_m=8$. This contribution is expected to inherit isocurvature perturbations from the inflationary power spectrum  eq.~\eqref{eq:isoFP} and, as we discuss in Section~\ref{ss:lateiso}, its abundance is constrained by CMB observations to be small, as is the case for $f_a\ll 10^{12}~\GeV$.

Soon after the time when $H = m_a$, the axion relaxes towards the minimum of its potential
in almost all patches, oscillating like matter, except those that contain a string or domain wall (which span several Hubble patches connecting the strings). As mentioned, this string/domain-wall network re-enters the horizon when the Universe's temperature is $T_{\rm PQ}<T_\star$
.  For $T_{\rm PQ}\ll T_\star$, the energy of the system at these times is dominated by that in domain walls rather than in strings. The strings and domain walls are expected to subsequently annihilate away in order-one Hubble times, producing a contribution to the axion relic abundance.

The dynamics of the string-wall system are highly nonlinear, and a reliable determination of the relic abundance that it produces would require dedicated simulations, which are likely to be  challenging given the large difference in scales between the Hubble rate $H_{PQ}$ when strings reenter the horizon and the other scales, $H_{PQ}\ll m_a\ll f_a$. Nevertheless, we can make a reasonable estimate of the axion relic abundance as follows.
\vspace{-1mm}
\begin{itemize}[leftmargin=0.2in] \setlength\itemsep{0.15em}
	\item  If $T_{\rm PQ}\gtrsim 150$ MeV, the strings and domain walls re-enter the horizon and decay while the axion mass has not yet saturated to its zero-temperature value. Assuming that the system instantaneously decays into non-relativistic axions, the resulting relic abundance is
\begin{equation}\label{eq:Omegahigh}
\frac{\Omega_a^{\rm dw}}{\Omega_{\rm dm}}=\frac{\rho_w(T_{\rm PQ})}{\rho_r(T_{\rm eq})}\frac{T_{\rm eq}}{T_{\rm PQ}}\left[\frac{g_{*}
(T_{\rm eq})}{g_{*}
(T_{\rm PQ})}\right]^\frac13\simeq 1.2\mathcal{A}\left[ \frac{f_a}{10^{10} \,{\rm GeV}}\right]\left[ \frac{10}{g_{*}(T_{\rm PQ})}\right]^\frac12\left[ \frac{150\, {\rm MeV}}{T_{\rm PQ}}\right]^5 \, ,
\end{equation}
where $\rho_w(T_{\rm PQ})\simeq 2\mathcal{A} \sigma(T_{\rm PQ}) H_{\rm PQ}$ is the energy in domain walls at $T=T_{\rm PQ}$ and $\sigma=\beta m_a(T_{\rm PQ}) f_a^2$ (with $\beta\simeq9$) their tension, $\rho_r$ the energy density in radiation and $T_{\rm eq}$ the temperature at matter-radiation equality (MRE). $\mathcal{A}$ is an order-one parameter measuring the domain wall area (in units of $1/H$) at the time when the string-wall system decays. As a result, for $T_{\rm PQ}\gtrsim 150$ MeV this contribution to the dark matter abundance is likely smaller than that from the axions that would be produced by the string scaling regime.

\item Conversely, if $T_{\rm PQ}\lesssim 150~{\rm MeV}$ the string-wall system decays when the mass has saturated to its zero temperature value and
\begin{equation}\label{eq:Omegalow}
\frac{\Omega_a^{\rm dw}}{\Omega_{\rm dm}}\simeq 1.8\mathcal{A}\left[ \frac{f_a}{10^8 \,{\rm GeV}}\right]\left[ \frac{10}{g_{*}
(T_{\rm PQ})}\right]^\frac12\left[ \frac{1\, {\rm MeV}}{T_{\rm PQ}}\right] \, .
\end{equation}
\end{itemize}

 This estimate assumes that the axions emitted are non-relativistic otherwise the contribution is more suppressed.
Note that as the network re-enters the horizon at lower temperatures $T_{PQ}$, the abundance $\Omega_a^{\rm dw}$ increases and correspondingly smaller $f_a$ is needed to reproduce the full dark matter abundance. 
Thus, the axion can make up the observed abundance for arbitrarily small $f_a$, with the  contribution in eq.~\eqref{eq:Omegalow} dominating that from the average-angle misalignment in eq.~\eqref{eq:Omegamis}. Eventually the required $f_a$ are in tension with the astrophysical bounds from star cooling, $f_a\gtrsim 10^8$ GeV \cite{Raffelt:2006cw,Viaux:2013lha,Ayala_2014,MillerBertolami:2014rka,Chang:2018rso,Carenza:2019pxu,Bar:2019ifz,Dessert:2022yqq}, which corresponds to $T_{\rm PQ}\sim 1$~MeV, around Big Bang Nucleosynthesis.

\begin{figure}[t!]
\centering
\includegraphics[width=0.7\linewidth]{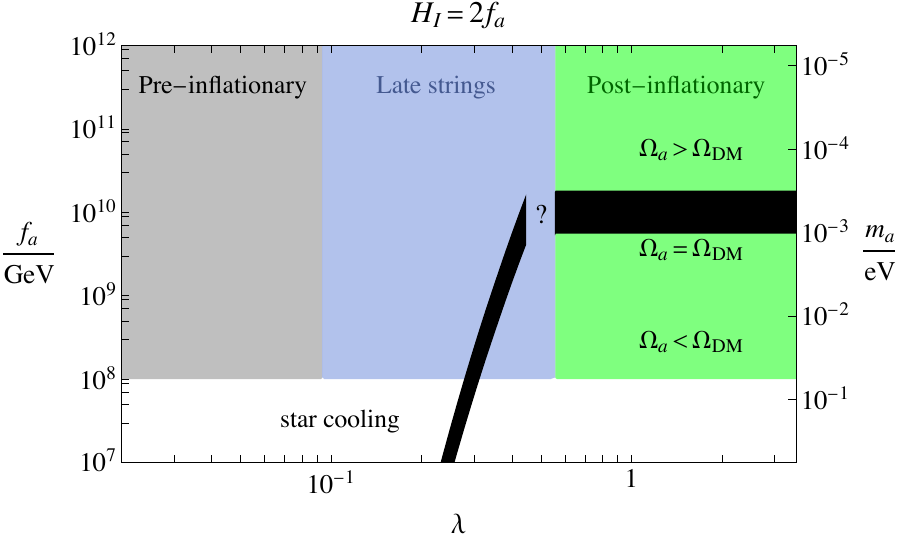}
\caption{\label{fig:summary}  Different axion scenarios, and the corresponding dark matter abundance, as a function of $\lambda$ and $f_a$ for $H_I=2f_a$ (assuming the string network re-enters the horizon when $F=0.35$, as in Figure~\ref{fig:landscape}). 
In the green region the string network 
re-enters the horizon when the temperature of the Universe $\gtrsim1$~GeV so that standard post-inflationary phenomenology is expected.
In blue the string-wall network re-enters the horizon and annihilates at temperatures $\lesssim1$~GeV, leading to novel phenomenology. In the grey region the network never forms within our observable horizon, as in the pre-inflationary scenario. 
The observed dark matter abundance is reproduced on the black line, which requires much smaller $f_a$ in the blue region compared to the standard post-inflationary scenario. The region $f_a\lesssim 10^8$ GeV is excluded by astrophysical constraints.
}
 \end{figure}

In Figure~\ref{fig:summary} we show the different behaviours and relic abundance as a function of $f_a$ and $\lambda$, keeping  fixed $H_I/f_a=2$. This corresponds to a slice through Figure~\ref{fig:landscape}
. 
The dark matter abundance is reproduced on the black line. In the green region this coincides with the axion post-inflationary scenario
where the abundance, even though currently uncertain, depends only on $f_a$. In the blue region where 
the annihilating domain walls dominate the abundance is reproduced for smaller $f_a$ up to the astrophysical bound, and follows eq.~\eqref{eq:Omegalow}. In the intermediate regime where the string and domain wall network re-enters the horizon soon after the QCD crossover, such that the abundance predicted from domain walls, eq.~\eqref{eq:Omegahigh}, is smaller than that expected from the scaling regime, we do not have a complete estimate for the dark matter abundance, so we leave a gap in the black line.

The dynamics of the late string-wall decay could also source stochastic GWs. Despite the uncertain dynamics, we can estimate the corresponding spectrum 
with the standard quadrupole approximation, similarly to theories with $N_W>1$, see e.g.~\cite{Gorghetto:2022ikz}. 
At $T=T_{PQ}$ the walls decay, emitting GW energy density at momentum $k\simeq H_{PQ}$ at a rate $d{\rho}_{gw}/dt \simeq G \rho_w^2/H_{PQ}$ for approximately one Hubble time. This leads to a peak amplitude today 
\begin{equation}\label{eq:Omegagw}
h^2\frac{d\Omega_{gw}}{d\log f}\simeq 5\cdot 10^{-18 }\mathcal{A}^2\left[ \frac{f_a}{10^8 \,{\rm GeV}}\right]\left[ \frac{1\, {\rm MeV}}{T_{\rm PQ}}\right]\simeq 10^{-18}\left[ \frac{\Omega_a^{\rm dw}}{\Omega_{\rm dm}}\right]^2\left[ \frac{1\, {\rm MeV}}{T_{\rm PQ}}\right] \, ,
\end{equation}
where 
we used eq.~\eqref{eq:Omegalow} and assumed $T_{PQ}\lesssim150$\,MeV. The corresponding frequency today is $f_{\rm peak}\simeq 2\cdot 10^{-11}\,\text{Hz}\,(T_{
\rm PQ}/1\,{\rm MeV})$. From eq.~\eqref{eq:Omegagw}, the GW spectrum is the largest for the smallest $T_{\rm PQ}$, and is a few orders of magnitude smaller the sensitivity of near-future detectors ($\Omega_{gw}\simeq 10^{-15}$) at the relevant frequency range ($f\gtrsim 10^{-9}$\,Hz). We stress however that eq.~\eqref{eq:Omegagw} is only an approximate parametric estimate, and the true GW emission could be substantially different.

\subsection{Axion spatial distribution and miniclusters} \label{ss:miniclusters}
Similarly to the standard post-inflationary scenario, the collapse of the string-wall system is expected to leave order-one inhomogeneities in the axion dark matter spatial distribution at small length scales. The typical comoving length $\lambda$ 
of the inhomogeneities is determined by the size of the domain walls at the time of collapse, approximately fixed by the inverse Hubble scale, i.e. $\lambda 
\simeq 1/(R_{PQ} H_{\rm PQ})$ (here $R_{\rm PQ}$ is the scale factor when $T=T_{PQ}$, and the network re-enters the horizon). This amounts to a contribution to the power spectrum of the axion overdensity field, $\Delta_\delta^2$ defined in eq.~\eqref{eq:delta2k} (specifically $\Delta_{\rm iso}^2$), that is peaked at the momentum $k/R_{\rm PQ}\simeq H_{\rm PQ}$, where it acquires an order-one value. For momenta $k/R_{\rm PQ}\ll H_{\rm PQ}$, this contribution to $\Delta_{\rm iso}^2$ is expected to fall $\propto C [k/(R_{PQ}H_{\rm PQ})]^3$ from causality arguments. Here $C$ is an order-one coefficient that could be in principle extracted from numerical simulations of the late string-wall decay.  Given the large uncertainties on the string-wall dynamics, the dark matter abundance and the spectrum, we refrain from giving a more detail description of the axion field, and briefly mention below some possible observational consequences of this dark matter substructure, see \cite{Baratella:2018pxi,Harigaya:2022pjd} for related discussions.

If the axion makes up a substantial fraction of the dark matter, the overdensities 
collapse at around matter-radiation equality (MRE), much before canonical structure formation takes place.  
A spherical overdensity with magnitude $\delta\gtrsim 1$ undergoes nonlinear collapse when the scale factor is $R\simeq R_{\rm eq}/\delta$, where $R_{\rm eq}$ is the scale factor at MRE. This collapse leads to the formation of a compact gravitationally bound object, known as axion minicluster, with average density  $\rho_b\simeq 140 \delta^3(1+\delta)\rho_c\simeq \mathcal{O}(100)\rho_c$, where $\rho_c$ is the average dark matter density at the time of collapse \cite{Hogan:1988mp,Kolb:1993zz,Kolb:1994fi,Zurek:2006sy}. Since the collapse happens around matter radiation equality, $\rho_b\simeq \, 100\,{\rm eV}^4$, about 8 orders of magnitude larger than average local dark matter density in the neighborhood of the Solar System.

The typical mass of the miniclusters is fixed by the dark matter mass contained in an order-one fluctuation, and reads
\begin{equation}\label{eq:minicluster}
    {M_{b}\simeq \frac{4\pi}{3} \left(\frac{2\pi}{x k_{\rm PQ}/R_{\rm PQ}}\right)^3 \bar{\rho}(R_{\rm eq})\simeq 24 M_\odot\left[ \frac{10}{g_{*}(T_{\rm PQ})}\right]^\frac12\left[ \frac{1\, {\rm MeV}}{T_{\rm PQ}}\right]^3x^{-3}}
\end{equation}
where $x\simeq\mathcal{O}(1)$ parameterizes the typical physical size of the fluctuations at $R=R_{\rm PQ}$ as $R_{\rm PQ}\lambda=2\pi/(xk_{\rm PQ}/R_{\rm PQ})$.  Owing to the larger size of the fluctuations (related to the larger Hubble distance at the time of string-wall decay), for $T_{\rm PQ}\ll 1 \,{\rm GeV}$ the miniclusters are significantly more massive than in the usual $N_W=1$ axion post-inflationary scenario, although their density is still of the order of the Universe's density at the time of collapse. 
Finally, the $k^3$ tail leads to the clustering of miniclusters into larger (more massive and less dense) compact halos at larger scales \cite{Fairbairn:2017sil,Enander:2017ogx,Vaquero:2018tib,Eggemeier:2019khm,Ellis:2020gtq,OHare:2021zrq}.

\subsection{Isocurvature constraints} \label{ss:lateiso}

As mentioned, the axion relic abundance from the late decay of domain walls is expected to have a white noise isocurvature spectrum $\Delta_{\rm iso}^2(k)=C [k/(R_{PQ}H_{\rm PQ})]^3$ at momenta $k/R_{PQ}\ll H_{\rm PQ}$, but are not expected to inherit the correlations from inflationary fluctuations, eqs.~\eqref{eq:isogauss} or \eqref{eq:isoFP}. Such white-noise isocurvature perturbations are constrained by CMB \cite{Planck:2018jri} and Lyman-$\alpha$  observations \cite{Croft:2000hs,Kobayashi:2017jcf,Murgia_2019,Irsic:2019iff,Feix:2019lpo,Rogers:2020ltq,Feix:2020txt}, which bound the fraction of isocurvature fluctuations $r_{\rm iso}\equiv \Delta_{\rm iso}^2(k_{\rm CMB})/\Delta_\mathcal{R}^2(k_{\rm CMB})$.  
In the late string-wall regime we have
\begin{equation}
{r^{1/2}_{\rm iso}=\frac{\Omega_a^{\rm dw}}{\Omega_{\rm dm}}\sqrt{\frac{C }{A_s}\frac{k_{\rm CMB}^3}{H_{\rm PQ}^3}}\simeq 4\cdot 10^{-4}~C^\frac12\mathcal{A}\left[ \frac{f_a}{10^8 \,{\rm GeV}}\right]\left[ \frac{10}{g_{*}(T_{\rm PQ})}\right]^\frac34\left[ \frac{1\, {\rm MeV}}{T_{\rm PQ}}\right]^\frac52} \, .
\end{equation}
Lyman-$\alpha$ observations require $r^{1/2}_{\rm iso}\lesssim 4\cdot 10^{-3}$ for a white-noise power spectrum \cite{Murgia_2019,Irsic:2019iff}, and provide a bound on the latest decay time in this scenario, which is slightly less stringent than the combination of those from dark matter overproduction and the astrophysical bound on $f_a$, see \cite{Redi:2022llj}.

In the late string regime, in addition to the axion relic abundance from domain walls there is a cleanly separated, sub-dominant, contribution from misalignment. As discussed in Section~\ref{ss:latestringrelic}, this is produced around the time of the QCD crossover and inherits any super-horizon correlations in $\cos\theta$ that remain at this time, and given that the axion is already massive when the network annihilates it is difficult to imagine that the correlations are destroyed at late times. Note however that the super-horizon correlations at the QCD crossover might be suppressed relative to those immediately at the end of inflation by overshoot, as discussed in Section~\ref{ss:isoover}. Given that the string network is expected to re-enter the horizon when $0.3 \lesssim F \lesssim 0.45$ is reached, which must happen long before CMB time, the corresponding $F$ at CMB scales is likely to be somewhat closer to $0.5$. Therefore, if eq.~\eqref{eq:Oversup} turns out to be the true dependence, the resulting suppression can be significant. Moreover, it is possible that the  small loops that are temporarily present even in an underdense network until they decay  might partly erase the super-horizon correlations prior to the QCD crossover. 
Considering the most pessimistic scenario that neither overshoot or the dynamics of loops erases the initial correlations, the resulting isocurvature power spectrum would be given by eq.~\eqref{eq:isoFP}, suppressed by the fraction squared of dark matter  that arises from misalignment, 
\begin{equation} \label{eq:isolate}
\Delta_{\rm iso}^2(k_{\rm CMB}) \sim \frac {\Omega_{\rm mis}^2} {\Omega_{\rm wall}^2} e^{-2 N_{\rm vis}/N_{11}} ~.
\end{equation}
Considering inflationary production,  we expect $N_{s}\simeq  N_{11}$ up to a numerical factor that we do not know reliably (e.g. assuming a critical $F=0.35$, $N_{s}\simeq 1.5 N_{11}$, c.f. eq.~\eqref{eq:N(F)} where $a_{11}b_{11}\simeq -0.65$). 
Consequently, in the late string regime with $25 \lesssim N_s\lesssim50$, we expect the suppression from the exponential factor in eq.~\eqref{eq:isolate} to be at least two orders of magnitude. (For overshoot production $N_{\rm vis}\ll N_{11}$ so the exponential does not lead to a suppression but, as mentioned in Section~\ref{ss:isoover}, an extra suppression from the overshooting dynamics is expected.) Meanwhile, from eq.~\eqref{eq:Omegamis} the abundance from misalignment is  suppressed for small $f_a$, with $\Omega_a^{\rm mis}/\Omega_a^{\rm dw}\simeq 10^{-5}$ for $f_a\simeq 10^{8}~\GeV$. Nevertheless, eq.~\eqref{eq:isolate} remains a potentially strong constraint given that $\Delta_{\rm iso}^2(k_{\rm CMB}) \lesssim 10^{-10}$ from Planck   data \cite{Planck:2018jri} (so even a fraction $10^{-4}$ of dark matter with with order one isocurvature perturbations would be excluded). At face value, this would require $f_a$ close to the current astro-physical limit for the late string regime bound to survive cosmological bounds, although we stress that our discussion is only tentative and future dedicated study would be useful.

\section{Discussion} \label{s:discussion}

In this work we explored a new regime of the minimal QCD axion that leads to novel phenomenology, including a modified dark matter abundance,  dominantly produced by domain walls annihilating after the QCD crossover, and larger mass axion miniclusters than in the standard post-inflationary scenario. Our discussion applies in particular to the simplest concrete realization of the post-inflationary axion, the KSVZ model with domain wall number $N_W=1$, and occurs over order-one ranges of $H_I/f_a$ and $\lambda$.

We focused on the scenario where the PQ symmetry is not restored by thermal effects, $T_{\rm max}\lesssim f_a$.
According to the common lore, for $H_I/(2\pi) \gtrsim f_a$ the PQ symmetry is restored during inflation while for $H_I/(2\pi)\lesssim f_a$ it is broken throughout the cosmological history of the universe.\footnote{We note that additional couplings between $\Phi$ and the inflaton could change these expectations \cite{Bao:2022hsg}.} 
Very different dynamics and phenomenology ought to apply to the two scenarios: in the first case a string network is expected to form realizing the axion post-inflationary scenario, while in the latter the axion pre-inflationary scenario would be realized. 

Clearly this picture must be incomplete and at least in the intermediate region $H_I/(2\pi) \simeq f_a$ deviations from post or pre-inflationary axion cosmology should be expected. Moreover, the boundaries of the two standard regimes are potentially important. As pointed out in Ref.~\cite{Lyth:1992tw}, the dynamics in fact depend on both $H_I/f_a$ and the quartic coupling $\lambda$ of the PQ complex field. Roughly the landscape is the following:
\begin{enumerate}
\item If the quartic  is sizable, say $\lambda \gtrsim  0.1$, and $H_I\gtrsim2 f_a$ inflationary fluctuations randomize the field  efficiently so that a string network forms soon after the end on inflation. 
Subsequently, the network rapidly approaches the scaling solution so that the details of the initial conditions are erased and the usual post-inflationary axion phenomenology is expected. 
\item For smaller $\lambda\lesssim 0.1$ the network formed soon after inflation is underdense (i.e. less than a string per Hubble patch) and will re-enter the horizon at later and later times during radiation domination until it re-enters after the axion starts to oscillate, i.e. at $T\lesssim 1$  GeV. Once this happens the phenomenology becomes very different, effectively realizing the meso-inflationary scenario of Ref.~\cite{Redi:2022llj,Harigaya:2022pjd} (but in a minimal theory without any additional interactions between e.g. $\Phi$ and the inflaton). The string network cannot reach the scaling regime and it effectively annihilates as soon as the perturbations re-enter the horizon. 
In this case, dark matter axions are dominantly produced by domain walls, allowing the full observed dark matter abundance to be obtained for smaller $f_a$ than in the axion post-inflationary scenario, down to the astrophysical bound $f_a\gtrsim 10^8$~GeV {(additionally, possible gravitational wave signals and primordial black holes resulting from similar dynamics have been considered in Refs.~\cite{Ge:2023rce,Ge:2023rrq})}. These features are  reminiscent of theories that reach scaling
early, with $N_W>1$ and a small energy bias between the vacua, in which case a domain wall network also survives beyond the QCD crossover \cite{Hiramatsu:2012sc}. However, $N_W>1$ theories require an energy bias in a narrow range and might not even be viable without reintroducing a strong CP problem \cite{Beyer:2022ywc}.
\item For even smaller $\lambda$ the PQ field is roughly homogeneous immediately after the end of inflation so that no topologically non-trivial configurations are initially produced. Contrary to {different} statements in the literature, we find that a string network can still form during reheating, for some values of $H_I/f_a$ and $\lambda$. During inflation indeed the PQ field can be displaced from the vacuum so that during reheating it can undergo oscillations (or `overshoot') over the top of its potential. In this case even small fluctuations allow the field to be randomized, and if this happens efficiently enough a string network forms, which can re-enter the horizon either at $T\gtrsim1$ GeV or at $T\lesssim1$ GeV (leading to the phenomenology in 1. and 2., respectively). If a network is not produced instead such theories are ruled out by isocurvature constraints.
\end{enumerate}

We believe that the qualitative behaviour outlined above is robust. However, obtaining precise quantitative predictions in terms of $H/f_a$ and $\lambda$ is challenging. We have approached this analytically by studying the evolution of $\Phi$ during and after inflation, which gives a rough estimate of the temperature when the initially underdense network is expected to reach one string per Hubble volume, and thus the scaling solution. Numerical simulations are needed to establish precisely the landscape in the plane $(H_I/f_a\,,\lambda)$, but these suffer from the challenges of large scale separations. In this paper we have also made a first step in this direction, and the results of our simulations suggest that the string density in an underdense network is indeed related to the spread of the radial mode over the top of its potential, determined by the quantity $F$ in eq.~\eqref{eq:Fdef}, and that overshoot can lead to a network. More work would be useful to better test the relation between $F$ and the string network density, e.g. starting from a wider range of initial conditions including initially more underdense networks. It would also be interesting to study the evolution and eventual destruction of a network that has not yet re-entered the horizon when the axion mass becomes cosmologically relevant. From preliminary investigation, this appears challenging in the simulations with $\simeq 2000^3$ lattice points that we have the computational power for, due to the multiple, widely separated, scales in the problem ($H_{PQ}\ll m_a \ll f_a$, where $H_{PQ}$ is the Hubble rate when strings reenter the horizon), but progress might be possible with very large grids.  

From a phenomenological point of view the main novelties arise when the network annihilates at temperatures $T\lesssim 1~\GeV$.  As discussed in Section~\ref{sec:iso}, in this regime misalignment gives a contribution to the axion relic abundance (sub-leading to that from domain walls discussed above in 2.), which is likely to induce isocurvature perturbations that are severely constrained by the CMB. A future precise determination of the resulting isocurvature power spectrum is therefore crucial to determine the viability of this regime. 
This also raises the question of how efficiently super-horizon correlations (as produced by inflation) are erased in the axion post-inflationary scenario when scaling is reached early, assuming the PQ symmetry is not restored by thermal effects. Although it is normally thought that such a regime is automatically safe from such constraints, as is the case for the thermal post-inflationary scenario, this relies crucially on the dynamics of strings in a way that, although expected, is yet to be confirmed in numerical simulations. Even a tiny contribution to isocurvature could exclude such models so it is important to determine how efficiently the large-scale correlations are erased by the string network. We hope to address this question with numerical simulation in the future.

The viability of this late string regime is of crucial importance because it opens the possibility of having QCD axion dark matter at low decay constants, down to those allowed by current astrophysical bounds $f_a~\simeq 10^8$ GeV, even in the most minimal KSVZ setup. This is exciting for direct detection experiments planning to search for QCD axions with large mass $m_a\lesssim 10^{-2}$~eV, such as IAXO~\cite{IAXO:2019mpb, IAXO:2020wwp} among others~\cite{Baryakhtar:2018doz,Mitridate:2020kly,BREAD:2021tpx}.  Additionally, for cosmological observations, it opens the possibility of the simultaneous existence of cold axions that form the bulk of the observed dark matter and a component of thermal axions (see e.g.~\cite{Baumann:2016wac,Ferreira:2018vjj,DEramo:2018vss,Notari:2022ffe,Bianchini:2023ubu} and references therein) detectable as an `effective additional neutrino species', $\Delta N_{\rm eff}$, via forthcoming CMB experiments such as CMB-S4~\cite{Abazajian:2019eic} and Simons Observatory~\cite{SimonsObservatory:2018koc} and in combination with Large Scale Structure surveys (e.g. Euclid~\cite{Brinckmann:2018owf}).

A final important caveat is that in the regime $H_I\gg f_a \lambda^{1/4}$ our results depend on the assumption that, prior to the visible e-folds of inflation, $\Phi$ acquired a value typical of its equilibrium distribution with the value of the radial mode of order eq.~\eqref{eq:vev}. If the radial mode instead had a much smaller value at the start of visible inflation, strings would form more easily than we predict because the radial mode would immediately spread over the top of its potential. Conversely, if the radial mode had much larger values at such times, say of order the Planck scale, the dynamics would again be different to our results. In the absence of a complete theory of inflation and the cosmology prior to this, the initial value of $\Phi$ will remain an important theoretical uncertainty in this part of parameter space.

{\small
	\subsubsection*{Acknowledgements}
	We wish to thank Andrea Tesi and Giovanni Villadoro for discussions. MR was supported by MIUR grant PRIN 2017FMJFMW.  
 EH acknowledges the UK Science and Technology Facilities Council for support through the Quantum Sensors for the Hidden Sector collaboration under the grant ST/T006145/1 and UK Research and Innovation Future Leader Fellowship MR/V024566/1. 
 The work of MG is supported by the Alexander von Humboldt foundation and has been partially funded by the Deutsche Forschungsgemeinschaft under Germany’s Excellence Strategy - EXC 2121 Quantum Universe - 390833306. The work of AN is supported by the grants PID2019-108122GB-C32 from the Spanish Ministry of Science and Innovation, Unit of Excellence Mar\'ia de Maeztu 2020-2023 of ICCUB (CEX2019-000918-M) and AGAUR2017-SGR-754. AN is grateful to the Physics Department of the University of Florence for the hospitality during the course of this work.
}

\appendix

\section{Approximate correlation functions}
\label{app:meanfield}

In this Appendix we present approximate formulas for the power spectrum of the PQ field at the end of inflation, see also the appendix of~\cite{Redi:2022zkt}. The derivations below give an alternative intuitive understanding for the correlation functions.

The power spectrum of a free scalar field $\phi$ with effective mass $M\lesssim H_I$ during inflation reads
\begin{equation}
P(k)= \frac{2\pi^2}{k^3}\Delta^2(k)= \frac {H_I^2} {2k^3} \left( \frac k {R H_I}\right)^{3-2\nu}\,,~~~~~~~~\nu= \sqrt{\frac 9 4- \frac {M^2}{H_I^2}}\approx \frac 3 2 -\frac {M^2}{3H_I^2}\,. \label{genericspectrum}
\end{equation}
From this we can compute the field's variance at the end of inflation,
\begin{equation}
\langle \phi^2(\mathbf{x}) \rangle = \frac {H_I^2}{(2\pi)^2} \int_{H_I R_e e^{-N_I}}^{H_I R_e} \frac {dk}k \left( \frac k {R_e H_I}\right)^{\frac {2M^2}{3H^2}}\approx 
\left\{\begin{array}{lcl} 
\displaystyle  \frac{3 H_I^4}{8\pi^2 M^2} & & N_I > \frac{H_I^2}{M^2}\,,\\
\displaystyle N_I \frac {H_I^2}{(2\pi)^2}  & & N_I < \frac{H_I^2}{M^2}\,\\
\end{array}
\right.\,,
\label{eq:vevAPPENDIX}
\end{equation}
where $R_e$ is the scale factor at the end of inflation and $N_I$ is the total number of e-folds of inflation. 
Eq.~\eqref{eq:vevAPPENDIX} shows that for small $M \ll H_I/\sqrt{N_I}$ it is a good approximation to neglect the potential and the variance undergoes a random walk, while for $M\simeq H_I/\sqrt{N_I}$ the variance reaches a constant value within the observable number $N_I$ of e-folds. The fact that $\langle \phi^2 \rangle$ becomes constant is due to the balance between the stochastic inflationary fluctuations and the classical force relaxing the field to the minimum. 
The convergence of the integral in the IR is due to the suppression of long modes, with momentum smaller than $H_IR_ee^{-N_I}$, that thus cease to contribute to $\langle \phi^2 \rangle$.  

Let us first consider the quartic potential $V=\lambda \phi^4/4$. The effective mass is $M^2= 3\lambda  \langle \phi^2 \rangle$ and, from eq.~(\ref{eq:vevAPPENDIX}), in the stationary regime we thus have
\begin{equation}
\langle \phi^2 \rangle\approx \frac{H_I^2}{\sqrt{8\pi^2 \lambda}} \, .
\end{equation} 
We can extend eq.~\eqref{eq:vevAPPENDIX} to several scalar fields, obtaining that the expectation value of $\phi_i\phi_j$ is
\begin{equation}
\langle \phi_i \phi_j \rangle= (M^2)^{-1}_{ij}\frac {3H_I^4}{8\pi^2} \, .
\label{eq:eq2fields}
\end{equation}
For the PQ field $V=\lambda/4 (\phi_1^2+\phi_2^2-f_a^2)^2$ so that,
\begin{equation}\label{eq:M2}
M^2 = \lambda
\begin{pmatrix}
3\langle\phi_1^2\rangle + \langle\phi_2^2 \rangle - f_a^2 & 2 \langle\phi_1 \phi_2  \rangle\\
2 \langle\phi_1 \phi_2\rangle & \langle\phi_1^2 \rangle+ 3 \langle \phi_2^2\rangle - f_a^2  \\
\end{pmatrix} \, .
\end{equation}
In the limit $H_I \gg \lambda^{1/4}f_a$, from eq.~\eqref{eq:eq2fields}
\begin{equation}
\lambda(3 \langle \phi_1^2 \rangle+ \langle \phi_2^2 \rangle)\langle \phi_1^2 \rangle=\frac {3H_I^4}{8\pi^2}\,,~~~~~\lambda(3 \langle \phi_2^2 \rangle+ \langle \phi_1^2 \rangle)\langle \phi_2^2 \rangle=\frac {3H_I^4}{8\pi^2}\,,
~~~~~\langle \phi_1 \phi_2 \rangle=0 \, ,
\end{equation}
with solution
\begin{equation}
\langle \phi_1^2  \rangle = \langle \phi_2^2  \rangle= \frac {\sqrt{3} H_I^2}{4\pi \sqrt{2\lambda}}
\label{eq:VEVapprox}
\end{equation}
By plugging eq.~\eqref{eq:VEVapprox} in eq.~\eqref{eq:M2}, and these in eq.~\eqref{genericspectrum}, we can also compute the power spectra,
\begin{equation}
\Delta^2_{\phi_1} (k)=\Delta^2_{\phi_2} (k)\approx \frac{H_I^2}{(2\pi)^2} \left( \frac k {R_e H_I} \right)^\gamma\,, ~~~~~\gamma= \sqrt{\frac{2\lambda}{3\pi^2}} \, .
\label{eq:Deltameanfield}
\end{equation}

The results above are equivalent to using the Hartree-Fock approximation, see \cite{Starobinsky:1994bd}.
The time scale, or the number of e-folds,  for the fields to reach the equilibrium distribution can be estimated by studying at the evolution of the expectation value $\langle \phi_i \phi_j \rangle$ during inflation. This is determined by the equation,
\begin{equation}
\frac {\partial}{\partial t} \langle \phi_i \phi_j\rangle=\frac {H_I^3}{4\pi^2}\delta_{ij}-\frac {2 M^2_{ik}}{3H_I}  \langle \phi_k \phi_j\rangle \, ,
\label{eq:meanfield}
\end{equation}
where the first term corresponds to the stochastic noise due to inflationary fluctuations while the second is the classical drift. Eq.~\eqref{eq:meanfield} is an approximation to the FP equation that assumes that the fields are gaussianly distributed, see~\cite{Starobinsky:1994bd} for the derivation. The time independent solution is given by eq.~\eqref{eq:VEVapprox}.
Indeed, taking $\langle \phi_i \phi_j \rangle=0$ and $\langle \phi_1^2\rangle=\langle \phi_2^2\rangle$, eq.~\eqref{eq:meanfield} simplifies to,
\begin{equation}
\frac {\partial}{\partial t} \langle \phi_1^2\rangle=\frac {H_I^3}{4\pi^2}- \lambda \frac {2 \langle \phi_1^2\rangle(4\langle \phi_1^2\rangle-f_a^2)}{3 H_I} 
\end{equation}
and at equilibrium one recovers eq.~\eqref{eq:VEVapprox}. 
The stationary solution is reached equating the random walk $\frac {H_I^2}{4\pi^2} N$ to the equilibrium value $\langle \phi_1^2 \rangle
$ in eq.~\eqref{eq:VEVapprox}, which gives a number of e-folds  $N\sim\sqrt{ \frac {3\pi^2}{2\lambda}}$.

\subsection{Isocurvature estimates}

We can also compute correlation functions relevant for isocurvature, as discussed in Section~\ref{sec:iso}. 
Assuming that the system has reached the asymptotic distribution we have,
\begin{equation}
\langle \cos \theta(x) \cos\theta(0)\rangle =  \langle \frac{\phi_1}{\sqrt{\phi_1^2 +\phi_2^2}}(x) \frac{\phi_1}{\sqrt{\phi_1^2 +\phi_2^2}}(0)\rangle \approx \frac 1 {\langle \phi_1^2+\phi_2^2\rangle}\langle \phi_1(x)\phi_1(0)\rangle \, ,
\end{equation}
In Fourier space,
\begin{equation}
\Delta^2_{\cos \theta}(k)= \frac{k^3}{2\pi^2} \int d^3 x e^{i k x} \langle \cos \theta(x) \cos\theta(0)\rangle= \sqrt{\frac {\lambda}{6\pi^2}}\left( \frac k {R_e H_I} \right)^{\sqrt{\frac{2\lambda}{3\pi^2}}} 
\end{equation}
where we used the expression for $\langle\phi^2_{1,2}\rangle$ in equilibrium in eq.~\eqref{eq:VEVapprox} and for the power spectrum for $\langle\phi_1(x)\phi_1(0)\rangle$. 
The suppression at the CMB scales is thus,
\begin{equation}
\label{eq:approxDeltacos}
\Delta^2_{\cos \theta}(k_{\rm CMB})\approx \sqrt{\frac {\lambda}{6\pi^2}} e^{-2N_{\rm vis}\sqrt{\frac{\lambda}{6\pi^2}}} 
\end{equation}
This has the same form of eq.~\eqref{eq:isoFP} with $N_{11}=\Gamma_{11}/H_I=\sqrt{6\pi^2/\lambda}\approx 8/\sqrt{\lambda}$.

\section{More on string production} \label{a:stringprod}

\begin{figure}[t]
	\centering
	\includegraphics[width=0.475\linewidth]{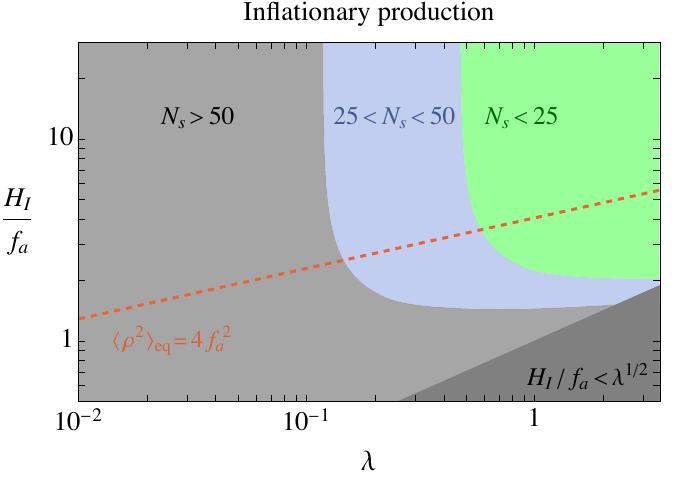} \quad
	\includegraphics[width=0.475\linewidth]{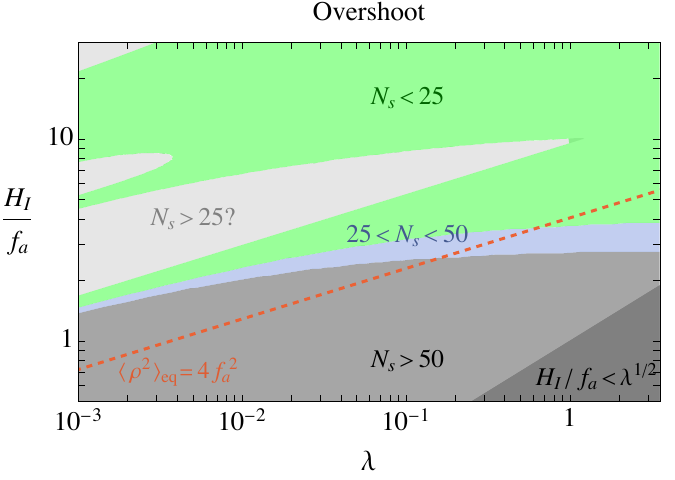}
	\caption{Number of e-folds of inflation $N_s$ that must re-enter the horizon for string network close to scaling to be reached, analogous to Figure~\ref{fig:Ninf} in the main text, except requiring a critical fraction $F=0.4$ rather than $0.35$. On the left, considering string production only from inflationary fluctuations and on the right only via the overshoot mechanism.}
\label{fig:Ninf2}
\end{figure}

In this Appendix we present a few additional results on our predictions of the number of e-folds of inflation that have to re-enter the horizon for a string network 
with $\xi\simeq 1$ to be reached (at which point the string scaling regime starts if $T_{PQ}\gtrsim1$~GeV, or the string-wall network annihilates if $T_{PQ}\lesssim1$~GeV). In particular, in Figure~\ref{fig:Ninf2} we show the same results as in Figure~\ref{fig:Ninf} in the main text, but with critical fraction $F=0.4$ rather than $0.35$, to give an indication of the uncertainty on our results (as shown in Figure~\ref{fig:sim1}, our results from numerical simulations are not precise enough to predict the critical $F$ to reach $\xi=1$ to better than the  range $0.3 \lesssim F \lesssim 0.45$; the critical value of $F$ might also vary depending on whether the string network is dominantly produced by inflationary fluctuations or the overshoot mechanism). In the left panel we show $N_s$ considering only inflationary production of strings. Given that in this case $N_s$ has only a logarithmic dependence on $F$, c.f. eq.~\eqref{eq:N(F)}, the shift in the contours of $N_s=25,~50$ are relatively mild, with changes of roughly less than order-one to the values of $\lambda$ and $H_I/f_a$ that, respectively, set the approximately vertical and horizontal contour regions. Meanwhile, in the right panel we show contours of $N_s$ considering only overshoot. The changes relative to Figure~\ref{fig:Ninf} are slightly larger in this case, with the structures on the left of the plot (corresponding to the radial mode starting in the middle of one of the `overshoot steps') getting somewhat larger. However, we stress that the exact details of the parameter space are especially uncertain in this regime given the unknown value of $\Phi_0$, so the additional uncertainty from $F$ is not a major issue.

\section{Further details of the numerical simulations} \label{app:sims}

\subsection{Description of simulations} \label{ss:sim_description}

We generate initial conditions for our simulations (made of $n^3$ lattice points) by the following procedure, equivalent to evolving the Langevin equation that the field solves during inflation, \cite{Lyth:1992tw}:
\begin{itemize}
\item We start with a homogeneous field $\Phi=\Phi_0$ on all lattice points. At this point, the entire simulation volume represents a  size $H_I^{-3}$ region $N=\log_2(n)$ e-folds before the end of inflation (with the value of the field coarse grained over $H_I^{-1}$).
\item To reproduce one doubling time of inflation (during which the initial $H_I^{-3}$ volume grows to $8 H_I^{-3}$) we split the homogeneous region into 8 patches. Inflationary fluctuations are accounted for by giving $\phi_1$ and $\phi_2$ (the canonically normalised real and imaginary parts of $\Phi$) in each of the new regions a `kick' drawn from a Gaussian distribution with variance $\log(2)H_I^2/(4\pi^2)$ and mean $0$. Finally, we evolve $\Phi$ in each of the regions according to the classical equations of motion (for a homogeneous field) for a time $\log(2)H_I^{-1}$. 
\item This procedure is repeated, with each homogeneous patch successively split into $8$, kicked and evolved until each grid point corresponds to an individual region of size $H_I^3$.
\end{itemize}
This generates an approximation to a realisation of the field configuration produced by inflation. As well as giving a distribution with a probability distribution of $\Phi$ that approximates the solution to the FP equation, such an approach captures higher order correlation functions, e.g. non-Gaussianities, that are present when the potential is non-negligible.

Subsequently the system is evolved forward in time from $t=H_I^{-1}$ using the classical equations of motion from the Lagrangian in eq.~\eqref{eq:L} discretised with a leapfrog algorithm that is accurate to second order in the timestep. We parameterize the cosmic time in simulations by $\log(m_\rho/H)$. As mentioned in the main text, the dynamics of the string network are only captured without significant systematics provided there is at least $\sim 1$ lattice point per string core (which is of size $m_\rho^{-1}$) and $HL\gtrsim 2$ (where $L$ is the physical size of the box). With the Lagrangian of eq.~\eqref{eq:L}, the string core size decreases in comoving coordinates $\propto R^{-1}$ (where $R$ is the scale factor). Consequently, to maximise the range in $m_\rho/H$ of such simulations, the string cores must initially contain $\sqrt{n}$ lattice points. In other words, $m_\rho \Delta =1/\sqrt{n}$ initially, where $n$ the number of lattice points. This is unfortunate for our purposes, because half of the e-folds worth of inflationary fluctuations are already inside the string cores at the start of the simulation, and we could only study $\frac{1}{2}\log n$ e-folds of inflation re-entering the horizon. Instead, we simulate the so-called fat string system, i.e. with $\lambda$ changing with time in such a way that $m_\rho\propto R^{-1}$, and we set the lattice spacing $\Delta$ as $m_\rho \Delta=1$; we start the simulations at $H=m_\rho$. In this way we can study the effects of almost the full generated $\log n$ e-folds worth of inflationary fluctuations reentering the horizon (although, as mentioned in the main text, the equations of motion differ from the physical system).

In practice, for initial conditions that give small $\xi$, statistical uncertainties on $\xi$ become large prior to the maximum time allowed by the systematic uncertainties discussed above (even with $\simeq 10$ repeated runs). Our data sets include sufficiently many simulations that the relative statistical uncertainty on $\xi$ is within $10\%$ (this is calculated via the standard error on the mean over repeated runs). We also only present results from timeshots $\log(m_\rho/H)\gtrsim 3$; prior to this the strings have a thickness comparable to the horizon size and are not cleanly defined objects (in the case of data sets with a large radial mode expectation value during inflation, we additionally only consider timeshots once the radial mode has settled down to the minimum of its potential in most of the simulation volume).

To produce the $x$ axis of Figure~\ref{fig:sim1}, we calculate the value of $F$, defined by eq.~\eqref{eq:Fdef}, as a function of the simulation time. As described in the main text, the relation between simulation time and $F$ depends on the values of $H_I/f_a$, $\lambda$ (used to generate the initial conditions), as well as the initial radial mode value $\rho_0$. One additional complication arises in the case of large initial field displacement (Figure~\ref{fig:sim1} right). 
In this regime, in fact, the value of the radial mode $\rho$ $N$ doubling times before the end of inflation is not unique in what corresponds to the comoving simulation volume. Instead, it has a distribution that occurs because the radial mode has evolved from the delta-function initial condition (peaked at $\rho_0$) to a non-trivial distribution during the preceding $11-N$ doubling times of inflation ($11$ is the total number of doubling times in the simulation, and appears given that our simulations with $2048^3$ lattice points lead to a total of $\log_2(2048)=11$ doublings). Thus, to compare to $\xi$ evaluated over the full simulation box (when $N$ doubling times of inflation have re-entered the horizon), we need to calculate $F$ taking into account this distribution. Although the results obtained for $F$ as a function of simulation time are not numerically very different from those obtained assuming a fixed value the radial mode $N$ doubling times before the end of inflation, this procedure results in $F$ being an increasing function of simulation time, as is expected physically. To produce the plot in Figure~\ref{fig:Ninf} we neglected this issue, and instead assumed that the radial mode value at $N$ e-folds before the end of inflation is unique.

In more detail, we calculate an average $F(N)$:
\begin{equation}\label{eq:Fav}
\bar{F}(N) = \int_{0}^{\infty} d\rho_i ~ P\left(\rho_i\right) F\left(N\right)|_{\rho=\rho_i} ~,
\end{equation}
where $P\left(\rho\right)=\int_0^{2\pi}d\theta~\rho P(\phi_1,\phi_2)|_{\phi_1^2+\phi_2^2=\rho^2}$ is the probability distribution of $\rho$ evaluated $N$ e-folds before the end of inflation (due to the preceding e-folds of inflation, in our case corresponding to $11-N$ doublings) and $F\left(N\right)|_{\rho=\rho_i}$ is the function defined by eq.~\eqref{eq:Fdef} setting $\rho=\rho_{i}$ $N$ e-folds before the end of inflation. For simulations with $\lambda=0$ in the initial conditions, $P$ is a Gaussian and we evaluate eq.~\eqref{eq:Fav} directly. These initial conditions lead to a field displaced by $\rho_0\gg f_a$ after inflation and should produce strings via overshoot, thus the issue discussed above is important to account for. For $\lambda\neq 0$ (and the values of $H_I/f_a$ used) instead the radial mode is not significantly displaced and we simply approximate that the probability distribution of $\rho$ is strongly peaked in the vacuum, which amounts to not carrying out the average in eq.~\eqref{eq:Fav}, and is a good approximation for all $(H_I/f_a,\lambda)$ plotted (additionally, it is numerically impractical to evaluate the required integrals for each timeshot for each set of initial conditions; we have however checked for individual data points that this approximation does not significantly affect the inferred $\bar{F}$).

\subsection{The string network} \label{ss:sim_xi}

\begin{figure}[t] 
	\centering
	\includegraphics[width=0.475\linewidth]{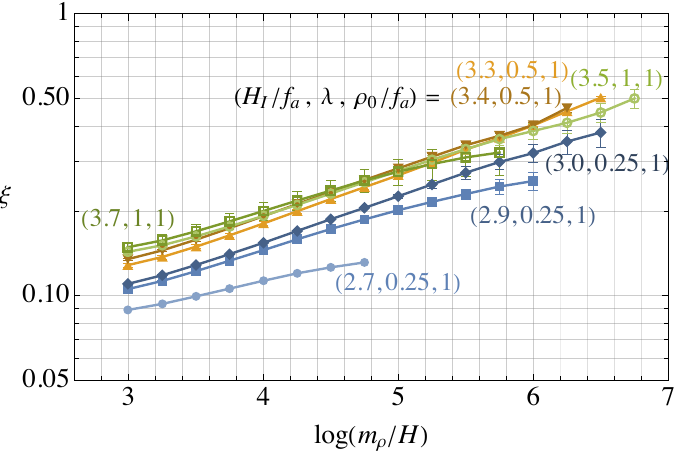}\quad
		\includegraphics[width=0.475\linewidth]{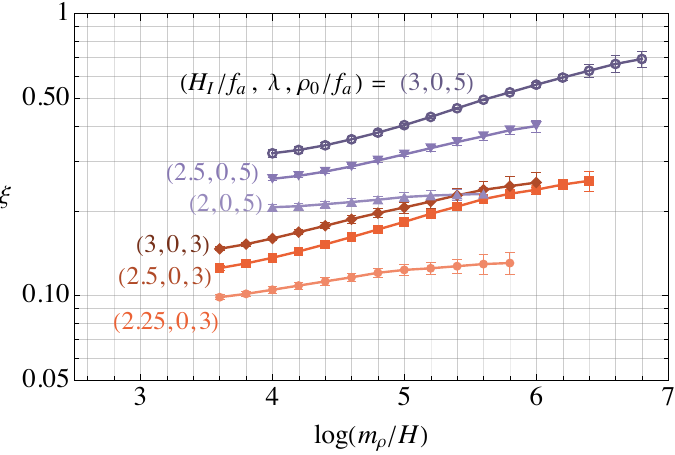}
	\caption{\label{fig:sim2}  The same simulation results as in Figure~\ref{fig:sim1}, but with the string density $\xi$ plotted as a function of simulation time, parameterized by $\log(m_\rho/H)$. In the left panel, results obtained from initial conditions such that strings form as a result of fluctuations directly over the top of the potential. Right panel, results with initial conditions such that strings form as a result of the radial mode overshooting the potential.}
\end{figure}

To clarify the properties of underdense string networks, we plot in Figure~\ref{fig:sim2} the string density $\xi$ as a function of simulation time (parameterised by $\log(m_\rho/H)$) rather than $F$. At a given simulation time the string density varies by a factor of $\sim 2$ among initial conditions with different $H_I/f_a$ and $\lambda$, and the rate of increase in $\xi$ also varies. It is therefore non-trivial that that the values of $\xi$ between different initial conditions coincide once $\xi$ is expressed as a function of $F$, as they do in Figure~\ref{fig:sim1}. It would however be interesting if simulations carried out on large grids could allow for underdense networks that have a greater range of $\xi$ at a given $\log(m_\rho/H)$ to test how well our proposed dependence on $F$ holds. As mentioned in the main text, the increase in $\xi$ with $\log(m_\rho/H)$ is exponential rather than linear. This is characteristic of the network being underdense compared to the critical $\xi$ such that string annihilation balances the rate at which strings reenter the horizon, as opposed to the logarithmic growth of $\xi$ on the attractor as a result of this critical point slowly moving with $\log(m_\rho/H)$.

\section{The effect of overshoot on the density power spectrum}\label{a:overshoot}

\begin{figure}[t]
	\centering
	\includegraphics[width=0.48\linewidth]{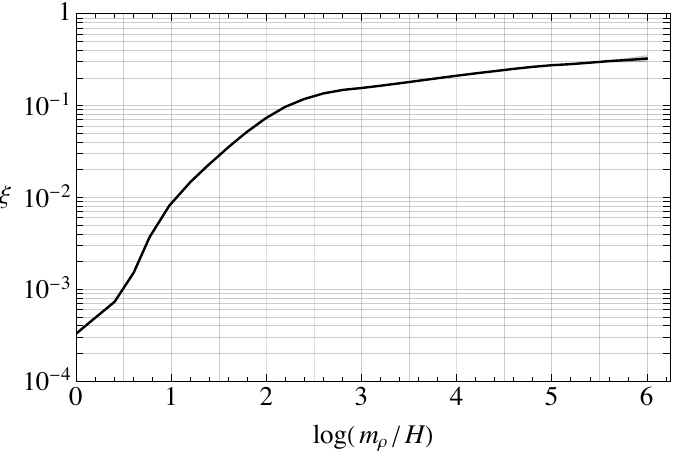} \quad
	\includegraphics[width=0.48\linewidth]{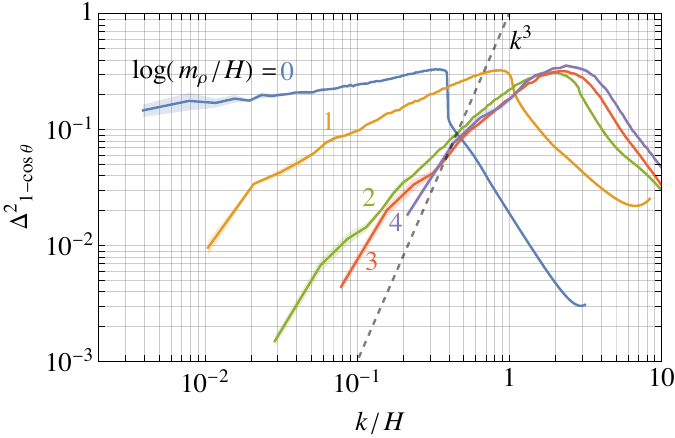}
	\caption{\label{fig:example_overshoot} 
  The evolution of the string density $\xi$ (left) and the power spectrum $\Delta^2_{1-\cos\theta}$ (right) in a numerical simulation of the string network, starting from initial conditions that lead to an initially under-dense network. We interpret the suppression of $\Delta^2_{1-\cos\theta}$ at large spatial scales, which occurs at early times in the simulation, as being due to overshoot randomising the phase of the complex scalar $\Phi$.
 }
\end{figure}

In this Appendix, we present a preliminary investigation of the effects of the relaxation of the PQ field to the minimum via overshoot on correlation functions at super-horizon scales, as discussed in Section~\ref{ss:isoover}. First, in Figure~\ref{fig:example_overshoot} we plot results for the string density $\xi$ and the power spectrum of $1-\cos(\theta)$, $\Delta^2_{1-\cos\theta}$, at different cosmological times (parameterized by $\log(m_\rho/H)$) for numerical simulations carried out similarly to described in Appendix~\ref{app:sims} (results are shown averaged over 5 simulation runs with different random realizations of the initial conditions). As initial conditions, we take $\phi_{1,2}$ to have a flat power spectrum with total variance $4 f_a^2$ centered around $\Phi=f_a/\sqrt{2}$, with the fluctuations cutoff at comoving momentum $k$ such that $k/(R_i f_a)\simeq 0.4$ where $R_i$ is the scale factor at the initial simulation time (when $H=f_a$). Such initial conditions lead to an initial under-dense string network with $\xi\ll 1$. We see that at the initial simulation time $\Delta^2_{1-\cos\theta}$ is almost flat (this is because a flat power spectrum for $\phi_1,\phi_2$ corresponds to a slighly tilted one for $1-\cos\theta$). However, a suppression a large scales quickly develops, already present by $\log(m_\rho/H)\simeq 2$, which we interpret as being due to the radial mode overshooting its potential (the string density at such times is still small, with most Hubble patches not containing a string). Note that the simulation is evolved in a background of radiation domination, so for the initial conditions we choose a sizable fraction of the radial mode overshoots, c.f. Figure~\ref{fig:phif}. The suppression seems to tend to the $k^3$ form expected if the axion field was completely randomised; a more detailed investigation would be useful. As an additional test, we have evolved the same initial conditions without the spatial gradients in the Hamiltonian. In such a system we again a similar suppression to $\Delta^2_{1-\cos\theta}$ on large scales, providing further evidence that this is due to overshoot.

\begin{figure}[t]
	\centering
	\includegraphics[width=0.48\linewidth]{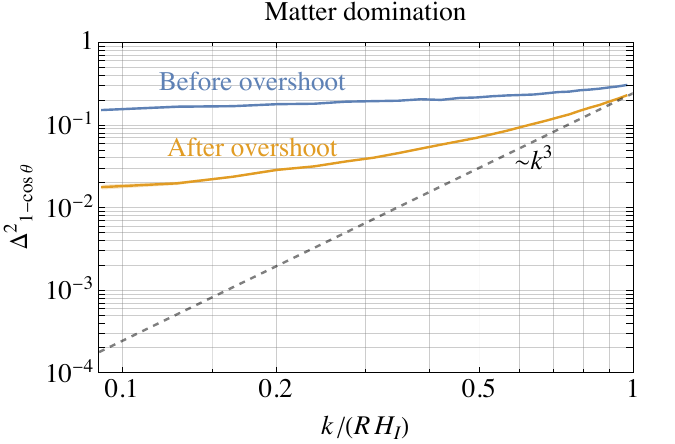} \quad
	\includegraphics[width=0.48\linewidth]{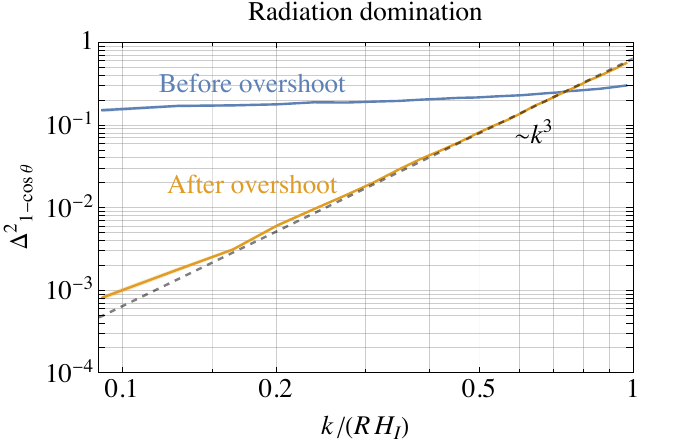}
	\caption{\label{fig:Povershoot}  The power spectrum of $1-\cos(\theta)$ for a realization of inflationary initial conditions consisting of $\phi_{1,2}$ having a flat power spectrum of (Gaussian) fluctuations of total variance of $(4f_a)^2$ (equivalent to a Hubble parameter during inflation of $H_I=4\pi f_a$) centred around $\Phi_0\simeq 4f_a$, as a function of comoving momentum $k/R$. Results are shown immediately after inflation (blue). Additionally, in orange we show results after mapping the radial mode to its vacuum using the overshoot function plotted in Figure~\ref{fig:phif} assuming matter domination (left) and radiation domination (right). Note that these are not results from a full numerical evolution of the theories equations of motion, and instead account only for overshoot and neglect spatial gradients. The suppression of $\Delta^2_{1-\cos\theta}$ at large scales is therefore due to the effects of overshoot. Numerically, in the case of matter domination the split of the radial mode on the two sides of its potential after overshooting is roughly $25\%-75\%$, and in this case a non-zero constant contribution to $\Delta^2_{1-\cos\theta}$ survives to large scales. In the case of radiation domination, overshoot is more efficient at randomising $\Phi$: there is a $49\%-51\%$ split of the radial mode between the two sides of the potential and $\Delta^2_{1-\cos\theta}$ is close to the $k^3$ white noise prediction for a completely randomized field.}
\end{figure}

As a further test, we generate a similar realisation of inflationary initial conditions with a flat initial power spectrum with total variance of $(4f_a)^2$ centered around $\Phi_0\simeq 4f_a$. Rather than evolving as in our field theory simulations, in this case we simply map the radial mode to its vacuum neglecting spatial gradients (and keeping the angular variable constant), i.e. using the step-like function plotted in Figure~\ref{fig:phif}. The form of $\Delta^2_{1-\cos\theta}$ before and after this mapping of the radial mode is plotted in Figure~\ref{fig:Povershoot}, assuming both the step-like function appropriate to matter domination and for radiation domination. A suppression of the super-horizon correlations by overshooting is evident. In the matter dominated case there is evidence that the {the destruction of large-distance correlations is not complete, i.e. $\Delta^2_{1-\cos\theta}$ tends to a non-zero constant as $k\to 0$, corresponding to the fact that the right hand side of  eq.~\eqref{eq:Oversup} is not zero if $F\neq 0.5$} (a less clear hint of a similar remaining correlation is also visible in the radiation dominated case). The numerical value that  $\Delta^2_{1-\cos\theta}$ tends to for $k\to 0$ is roughly compatible with eq.~\eqref{eq:Oversup}, given that $F\simeq 0.25,~0.49$ for matter domination and radiation domination respectively.

\pagestyle{plain}
\bibliographystyle{jhep}
\small
\bibliography{biblio}

\end{document}